\documentclass[a4paper,onecolumn,11pt,accepted=2021-11-17]{quantumarticle}
\pdfoutput=1
\usepackage[utf8]{inputenc}
\usepackage[english]{babel}
\usepackage[T1]{fontenc}
\usepackage{amsmath}
\usepackage{amssymb}
\usepackage{hyperref}
\usepackage{bm}% bold math

\newcommand{\beqar}{\begin{eqnarray}}
\newcommand{\eeqar}{\end{eqnarray}}

\newcommand{\bcen}{\begin{center}}
\newcommand{\ecen}{\end{center}}

\newcommand{\eps}{\varepsilon}
\newcommand{\lam}{\lambda}
\newcommand{\bra}[1]{\left< #1 \right|}
\newcommand{\ket}[1]{\left| #1 \right>}

\newcommand{\tr}{\mathrm{tr}}

\newcommand{\f}[2]{\frac{#1}{#2}}
\renewcommand{\b}[1]{\left({#1}\right)}
\renewcommand{\v}[1]{\vec{#1}}
\newcommand{\pd}[2]{\frac {\partial #1}{\partial #2}}

\renewcommand{\sb}[1]{\left[{#1}\right]}
\newcommand{\mean}[1]{\langle {#1} \rangle}

\newcommand{\ra}{\rightarrow}

\usepackage{tikz}
\usepackage{lipsum}
\usepackage[sort,numbers]{natbib}

\begin{document}

\title{{Quantum thermo-dynamical construction for driven open quantum systems}}

\author{Roie Dann}
\email{roie.dann@mail.huji.ac.il}
%\altaffiliation[Also at ]{The Institute of Chemistry, The Hebrew University of Jerusalem, Jerusalem 9190401, Israel.}%Lines break automatically or can be forced with \\
\affiliation{The Institute of Chemistry, The Hebrew University of Jerusalem, Jerusalem 9190401, Israel}%
\orcid{0-0002-8883-790X}
\author{Ronnie Kosloff}%
\email{kosloff1948@gmail.com}
\affiliation{The Institute of Chemistry, The Hebrew University of Jerusalem, Jerusalem 9190401, Israel}
\orcid{0000-0001-6201-2523}

\maketitle

\begin{abstract}
Quantum dynamics of driven open systems should be compatible with both quantum mechanic and thermodynamic principles. By formulating the thermodynamic principles in terms of a set of postulates we obtain a thermodynamically consistent master equation. 
Following an axiomatic approach, we base the analysis on an autonomous description, incorporating the drive as a large transient control quantum system. In the appropriate physical limit, we derive the semi-classical description, where the control is incorporated as a time-dependent term in the system Hamiltonian. The transition to the semi-classical description reflects the conservation of global coherence and highlights the crucial role of coherence in the initial control state.
We demonstrate the theory by analyzing  a qubit controlled by a single bosonic mode in a coherent state.
%This construction is employed to obtain the master equation of a driven quantum system with a time-depenedent Hamiltonian.
\end{abstract}

\section{Introduction}

Any realizable quantum system interacts with its environment to some extent. As a consequence,  accurate modeling and simulation of quantum dynamics  requires equations of motion which incorporate the  environmental influence. Typical experiments include control of the quantum entity by time-dependent fields. The simultaneous influence of the external field and environment are inevitably interrelated, i.e., the control modifies the system-environment interaction and vice versa. 
Modeling the dynamics in such a scenario is commonly carried out by a master equation with a time-dependent Hamiltonian and dissipative components.  
In order to accurately describe reality, such equations of motion must reflect the fundamental physical principles. Specifically, the dynamical equations of  the reduced system must comply with both thermodynamical and quantum mechanical principles. Our current objective is to derive a consistent dynamical equation of motion for a driven open quantum system.

The standard approach to derive the equation of motion for the reduced quantum system is based on embedding the system in an extended Hilbert space which includes the environment  \cite{kraus71}. Within this framework, the quantum dynamics are described by a unitary transformation, generated by a global Hamiltonian.  Reduced description of the system is then obtained by tracing over the {environmental} degrees of freedom.

Reduced dynamical equations have been obtained primarily by two methodologies. The first, termed microscopic derivation, starts from the global embedded description and employs a series of approximations. Such derivations are typically based on a second order perturbation theory in the system-environment coupling and memory-less dynamics \cite{davies1974markovian,breuer2002theory}. Generally, microscopic derivations do not  guarantee consistency with thermodynamics {or even positive semi-definiteness of the system state}. It is therefore customary, prior to the derivation, to check the validity of the dynamical equation with respect to physical principles. {In addition, the range of validity of the derived the master equation is limited by the set of approximations employed in the derivation. There are a number of proposed constructions, deriving the reduced dynamics of driven open quantum systems. For slow driving a quantum adiabatic approach has been employed to obtain the master equation \cite{albash2012quantum,yamaguchi2017markovian}. In addition, under periodic driving, a derivation based on a Floquet analysis gives the so-called Floquet master equation \cite{alicki2012periodically,szczygielski2013markovian}. We have previously, employed a general approach for driven systems based on the Davies construction which generalizes these two cases. This derivation leads} the Non-Adiabatic Master Equation (NAME) \cite{dann2018time}.

An alternative methodology to obtain the reduced dynamical equation of motion relies on a thematic approach. In this framework,  the general physical principles and symmetries of the dynamics are expressed  in terms a set of mathematical restrictions (axioms). The basic principles then allow to formulate the general structure of the equations of motion. 

Based on this embedding principle and the thematic approach, the seminal work of Gorini, Kossakowski, Lindblad and Sudarshan (GKLS) obtained the general form of the Markovian master equation \cite{gorini1976completely,lindblad1976generators}. This construction guarantees consistency with the probability interpretation of quantum mechanics, nevertheless, in certain cases it can violate  thermodynamic principles \cite{kohen1997phase,levy2014local}. We have recently addressed this flaw for a system represented by a static Hamiltonian. The remedy  was to introduce an additional constraint on the system-environment coupling. Namely, we assumed that the interaction satisfies strict energy conservation between the system and environment. Such a constraint is a manifestation of an additional symmetry which implies that the dynamical map is covariant with respect to the free propagation \cite{holevo1993note}. This restriction leads to the general structure of the master equation which complies with the basic laws of thermodynamics \cite{dann2020open} (without relying on further assumptions). In the present study, we extend this methodology to obtain the dynamical equation for a driven open quantum system.  The resulting master equation is of the GKLS form, where the eigenoperators of the free propagator constitute the Lindblad jump operators. The resulting dynamical structure can be employed to validate  master equations obtained by alternative methods. 

The key idea in addressing the time-dependent scenario is to describe the time-dependent drive as an additional quantum  system, which we termed the {control system}. In such a description, the explicit time dependence is replaced by an initial non-stationary state of the {control}. This procedure can be viewed as an additional embedding of the driven system within a larger Hilbert space, with dynamics generated by a static Hamiltonian. As will be shown, under such dynamics the total coherence is conserved. This property {highlights} the role of coherence as a resource in the embedded description \cite{lostaglio2015quantum,streltsov2017colloquium}. That is, coherence that originated in the initial non-stationary state of the {control} is transferred to the primary-system. 
From the reduced perspective of the system,
the coherence transfer 
can be viewed as an effective time-dependent driving.  

We define two possible treatments of the time-dependent control: the {\emph{autonomous}} and {\emph{semi-classical}} descriptions. 
In the {\emph{autonomous}} approach the dynamics are generated by a time-independent Hamilitonian. As a consequence the evolution is fully determined by  the initial state. In the limit of a macroscopic {control} with respect to the system, the control can be effectively replaced by time-dependent parameters, which defines the {\emph{semi-classical}} description. 
In this paper, we analyse the relation between the two descriptions and present a systematic procedure to translate from the autonomous to the semi-classical description. The relation between the two frameworks serves as a crucial tool in our analysis.
 We will demonstrate the autonomous-semi-classical transition by analyzing an explicit example, of a periodically driven qubit.
 
{The outcome of our analysis is a semi-classical master equation which is derived from an axiomatic approach and is thermodynamically consistent. The structure of this equation justifies and extends the validity regime of the derived master equations, the adiabatic, Floquet and NAME. In addition, the axiomatic construction serves as a bridge between a dynamical description and ideas from quantum thermodynamic resource theory \cite{janzing2000thermodynamic,brandao2013resource,horodecki2013fundamental}}.

\section{Framework and thermodynamically motivated postulates}
\label{sec:framework_postulates}

\begin{figure}
\centering
\includegraphics[width=8.cm]{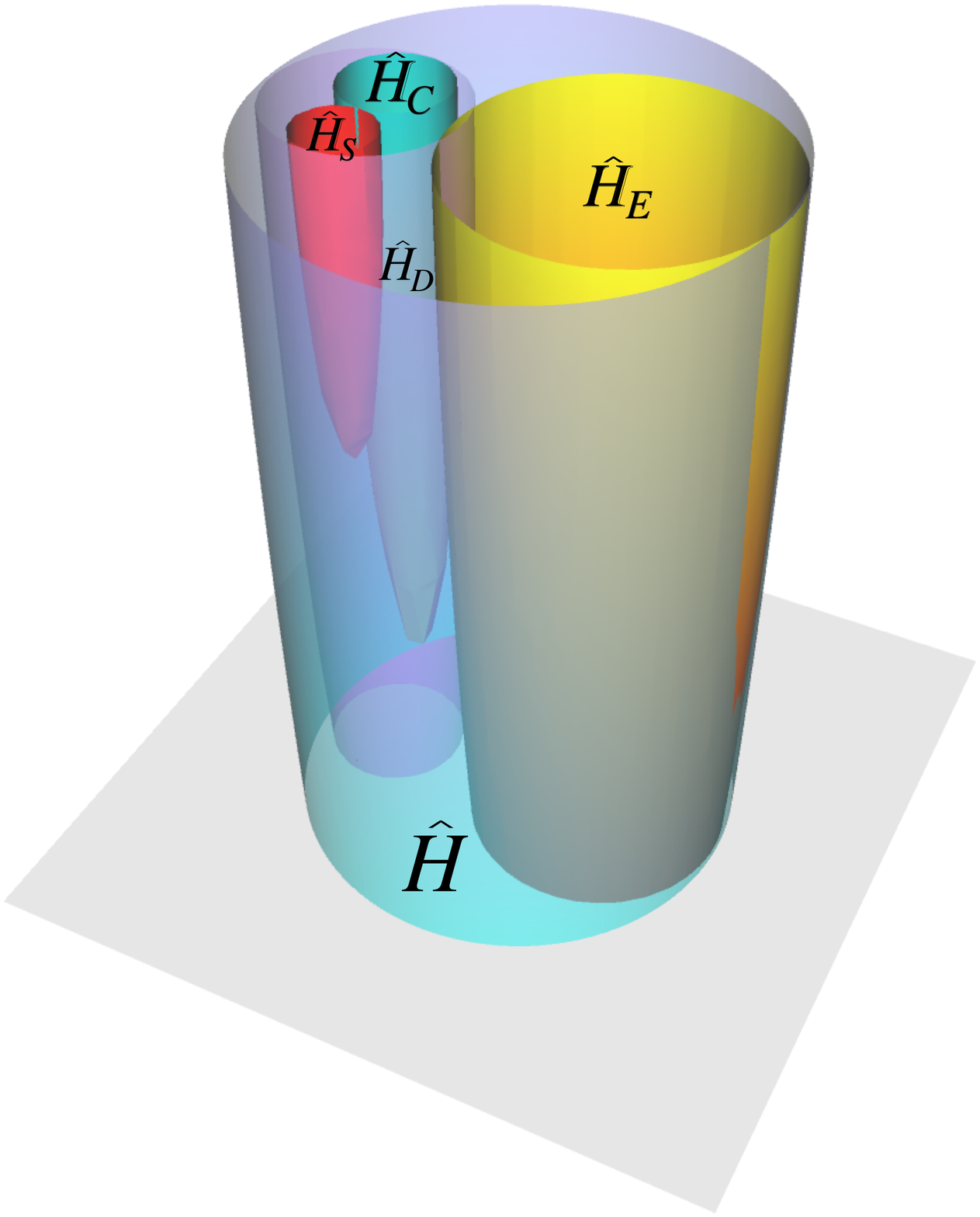}
\caption{
{An illustration of the embedded setup (Matryoshka): the decomposition of the total Hamiltonian $\hat{H}$, Eq. \eqref{eq:compsite_Hamiltonian} into the device (transparent blue tube) and environment  (yellow tube). The device is composed of a primary-system (red) and control (pale blue). Both the primary-system and control are coupled to the environment by the coupling term $\hat{H}_{DE}$, Eq. \eqref{eq:compsite_Hamiltonian}, while the primary-system and control are coupled by an internal coupling $\hat{H}_{SC}$, Eq. \eqref{eq:autonomous_hamil}. The total Hilbert space is decomposed into ${\mathfrak{D}}\otimes\mathfrak{E}$, where $\mathfrak{D}$ and $\mathfrak{E}$ are the Hilbert spaces of the device and environment, correspondingly. The device's Hilbert space can be decomposed to the Hilbert space of the  primary-system $\mathfrak{S}$ and control $\mathfrak{C}$. We implicitly assume that the Hilbert space of environment is much larger then the Hilbert space of the device, and the system's Hilbert space is the smallest.
}}
\label{fig:1}
\end{figure}

To set the stage, we begin with a complete quantum description of  all the thermodynamic constituents. By including the control apparatus within the autonomous quantum description, we identify the ``device system'' which is composed of a {\emph{primary}}-system  and the {control}, see Fig. \ref{fig:1}. The Hamiltonian of such a system is of the form 
\begin{equation}
    \hat{H}_D = \hat{H}_S +\hat{H}_C +\hat{H}_{SC}~~,
    \label{eq:autonomous_hamil}
\end{equation}
where $\hat{H}_S$ and $\hat{H}_C$ are the bare primary-system and {control} Hamiltonians, {which domain is the system and control Hilbert space $\mathfrak S$ and $\mathfrak{C}$, and  $\hat{H}_{SC}$ is the primary-system-control interaction term, which domain is the device Hilbert space $\mathfrak{D}$}. {The inclusion of the control system within the device is the key conceptual step which will allow bridging the gap between the autonomous and semi-classical descriptions.}
Complementing the internal interaction $\hat{H}_{SC}$, the device system also interacts with an external environment. Hence, the complete system, including {the primary-system, control and environment} is represented by 
\begin{equation}
    \hat{H} = \hat{H}_D+\hat{H}_{DE}+\hat{H}_E~~,
    \label{eq:compsite_Hamiltonian}
\end{equation}
where $\hat{H}_E\in{\mathfrak{E}}$ is the bare environment Hamiltonian, and the $\hat{H}_{DE}$ is the system-environment interaction term. 

{ In the present context, the term ``autonomous'' alludes to the fact that the underlying framework, Eqs. \eqref{eq:autonomous_hamil} and \eqref{eq:compsite_Hamiltonian}, incorporates all the relevant subsystems and describes them  as quantized systems, which are represented by time-independent Hamiltonian terms. As a consequence, the only source of time-dependence originates from ``intrinsic'' properties taken into account in Hamiltonian \eqref{eq:compsite_Hamiltonian} and the initial global state.}

%\tg{The theory of open quantum systems
%describes
%the dynamics of a quantum system coupled to an external environment. This theory is founded on the dogmatic approach of the `church of the larger Hilbert space', deducing the reduced quantum dynamics from an autonomous quantum description of the environment, Eq. \eqref{eq:compsite_Hamiltonian}.} 
When the initial device-environment state is uncorrelated the reduced dynamics are given by a completely-positive-trace-preserving (CPTP) map:
\begin{equation}
\hat{\rho}_D\b t={\Lambda}_t\sb{\hat{\rho}_D\b 0}\\=\text{tr}_E\b{\hat{U}\b{t,0}\hat{\rho}_D\b 0\otimes \hat{\rho}_E\b 0\hat{U}^\dagger\b{t,0}}~~, 
\label{eq:cal_Lambda_t}
\end{equation}
where $\hat{\rho}_E\b 0$ {and $\hat{\rho}_D\b 0$ are} the initial states of the environment {and device,}  and $\text{tr}_i$ signifies a partial trace over the $i$ system. {Throughout this study we label states, operators and superoperators of a subsystem with a corresponding subscript, for example primary system operators are designated as $\hat{X}_S$.} The joint system dynamics is generated by Eq. \eqref{eq:compsite_Hamiltonian}, and are governed by the `{\emph{autonomous propagator}}'  $\hat{U}\b{t,0}=e^{-i\hat{H}t/\hbar}$. 
An exact solution of Eq. \eqref{eq:cal_Lambda_t} is intractable, as the environment has a vast number of degrees of freedoms. Therefore, the approach commonly adopted starts from {a complete} description and derives the reduced system dynamics by conducting a set of approximations \cite{nakajima1958quantum,zwanzig1960ensemble,davies1974markovian,alicki1977markov,diosi1993calderia,lidar2001completely,daffer2004depolarizing,shabani2005completely,maniscalco2006non,breuer2008quantum,whitney2008staying,piilo2008non,alicki2012periodically,szczygielski2013markovian,majenz2013coarse,albash2012quantum,muller2017deriving,smirnov2018theory,dann2018time,mccauley2019completely,mozgunov2020completely,nathan2020universal}. As a consequence, the equation of motion obtained, the master equation, has a restricted validity regime, determined by the approximations involved. 

In the present paper, we follow an alternative methodology \cite{dann2020open}, introducing  four thermodynamically motivated postulated which complement the fundamental postulates of quantum mechanics.
Relying on these postulates, we proved in Ref. \cite{dann2020open} that the reduced system dynamics must be of a limited form (with no additional approximation). The emergent structure of the master equation is consistent with the weak coupling limit Davies construction \cite{davies1974markovian} and scattering or collision models
\cite{dumcke1985low,rodrigues2019thermodynamics}.

The four postulates are:
\begin{enumerate}
    \item The dynamical map $\Lambda_t$ is Markovian, satisfying the semi-group property: $\Lambda_{t}=\Lambda_s\Lambda_{t-s}$ for any $t,s\in \mathbb{R}$.
    \item The environment is initially in a stationary state with respect to the environment's free Hamiltonian $\hat{H}_E$. In the thermodynamic limit, we can extend this condition to all times.
    \item The dynamical map has a unique fixed point.
    \item  The composite system satisfies {\emph {strict energy conservation}} between the device and environment: $\sb{\hat{H}_{DE},\hat{H}_D+\hat{H}_E}=0$.
\end{enumerate}

{Postulate 1 states that any (invertible) dynamical map up to time $t$ can be written as the composition of two parts, taking any intermediate time $s$ as a stepping stone between $0$ and $t$.} The statement implies that the memory of the system's past state does not influence the present map, i.e., the dynamics are of Markovian nature. This is clearly an idealization, since, the coupling between system and environment leads to classical and quantum correlations between the components of the composite system. Information on the system's past state is encoded within these global correlations, which do effect the evolution of the system. Moreover, the complete dynamics (Eq. \eqref{eq:compsite_Hamiltonian}) are unitary, thus, the dynamics are reversible and in principle no information is lost. Nevertheless, once the information is encoded in global correlations (including many degrees of freedom), it becomes inaccessible and is practically lost. Consequently, when the environment correlations decay sufficiently fast the system dynamics are essentially Markovian. This idealization is valid within a coarse-graining time, which is of the order of the typical time associated with the decay of the environment's inner correlations. {Note that the concept of Markovianity in the quantum context has a number of possible definitions. A common choice identifies Markovianity with the CP divisibility of the dynamical map \cite{rivas2014quantum,breuer2016colloquium,li2018concepts}}. This serves as a weaker version of postulate 1, nevertheless, it is also sufficient for the derivation of the driven master equation. 
%The considered Markovian property implies that the composite state must remain separable throughout the evolution. %\cite{lindblad1996existence,lindblad1998brownian}.
%which enables a {reduced description} of the system.  \tb{discuss more}

The second postulate limits the description to environments which are diagonal in the eigenbasis of $\hat{H}_E$. These include a single or multiple thermal baths with arbitrary temperatures, all coupled simultaneously to the system. Conversely, squeezed baths are not included in the present description. The idealization, assuming the environment remains stationary, is motivated by the fact that the environment is much larger than the system. As a result, its state is only negligibly altered by the interaction with the system.

Postulate 3 is motivated by the $0$-law of thermodynamics, which infers that thermalization leads the system to a unique fixed point \cite{alicki1976detailed,alicki2018introduction}. The fourth postulate, strict energy  conservation, implies that energy is not accumulated within the interface between the system and environment \cite{janzing2000thermodynamic}. This is naturally satisfied in the thermodynamic limit, where the interface energy is negligible with respect to the bulk energy. Moreover, in the weak coupling regime and low density limit, postulate 4 is satisfied as an asymptotic limit. For a critical analysis on the strict energy conservation condition and its implications see Ref. \cite{dann2020open} Sec. III.

\section{Autonomous open system dynamics from thermodynamic principles}
\label{sec:time_independent}
The route to the autonomous description of the open systems starts from the complete description Eq. \eqref{eq:compsite_Hamiltonian}, which introduces a partitions between the device and the environment and partitions the device into the primary-system and control. {Building upon this partition} we employ the four thermodynamic postulates of Sec. \ref{sec:framework_postulates}  to determine the general structure of the master equation. {The presented construction of the autonomous reduced dynamics is based on the analysis presented in Ref. \cite{dann2020open}. Here, we briefly describe the key steps of the procedure, and derive the fixed point of the dynamical map.  The obtained master equation will serve as a building block to derive the semi-classical dynamical equation in Sec. \ref{sec:open_driven_system}.} The procedure {employs} a spectral analysis of the dynamical maps and their generators {and relies on their symmetry properties} . 

The first postulate implies that the dynamical map can be expressed in terms of the dynamical semi-group generator $\cal L$: $\Lambda_t=e^{{\cal{L}}t}$. The generator  can be decomposed into two terms corresponding to a unitary contribution and a dissipative part
\begin{equation}
{\cal{L}}=-i{\cal H}+{\cal{D}}~~.
\label{eq:L_decompos}
\end{equation}
The unitary part is given by ${\cal H}\sb{\bullet}=\hbar^{-1}\sb{\hat{H}+\hat{H}_{LS},\bullet}$, where $\hat{H}_{LS}$ is the {Lamb-shift} term and $\cal D$ is of the  GKLS form.

An additional major restriction emerges from the commutativity properties of the dynamical map, {which imply a time-translation symmetry of the dynamical map with respect to the free dynamics.} %\tg{which imply additional symmetries according to Wigner \cite{wigner1967symmetries}.} 
Specifically, when strict energy conservation (postulate 4) is satisfied and the environment is initially in a stationary state (postulate 2), the open system dynamical map commutes with the dynamical map of the free dynamics ${\cal U}_D$ (governing the {device} dynamics of) \cite{marvian2014extending,marvian2012symmetry,lostaglio2015description,dann2020open}
\begin{equation}
    {\cal{U}}_D\sb{\Lambda\sb{\hat{\rho}_D}}={\Lambda\sb{{\cal{U}}_D\sb{\hat{\rho}_D}}}~~.
    \label{eq:maps_commute}
\end{equation}
Here, ${\cal U}_D=e^{-i{\cal{H}}_D t}$, where ${\cal H}_D$ is the generator of the unitary map ${\cal H}_D\sb{{\hat{\rho}_D\b t}}={\hbar^{-1}}\sb{\hat{H}_D,\hat{\rho}_D\b t}$.
The {\emph{commutativity property}}{, denoted as time-translation symmetry,} imposes a strict structure on the dynamical map $\Lambda$. {Namely, a} spectral {analysis infers}
 that the two dynamical maps share a common basis of eigenoperators. %\tb{define the eigenoperators here?}

The eigenoperators of ${\cal U}_D$ can be classified into two types, invariant and non-invariant operators, both reside within the device's Hilbert space $\mathfrak{D}$. The invariant operators are composed of the energy projection operators $\{\hat{\Pi}_j=\ket{\psi_j}\bra{\psi_j}\}$ and have a vanishing eigenvalue, whereas the non-invariant are the transition operators $\{\hat{G}_{nm}=\ket{\psi_n}\bra{\psi_m}\}$ with the Bohr frequencies of $\hat{H}_D$, $\{\omega_{nm}=\b{\eps_m-\eps_n}/\hbar\}$ as eigenvalues.  When the Bohr frequencies are non-degenerate, i.e. $\omega_{nm}\neq\omega_{kl}$ for different $n,m,k$ and $l$, $\{\hat{G}_{nm}\}$ also constitute eigenoperators of open system dynamical map $\Lambda$. For the case of degenerate transition operators, the eigenoperators can be a general combination of the degenerate operators. For simplicity, in the following analysis we assume that the Bohr frequencies are non-degenerate.

Under the four thermodynamic postulates, the spectral properties of the dynamical map impose the following structure for the dissipator
\begin{equation}
    {\cal{D}}\sb{\bullet} = \sum_{{n,m}=1}^{N}\gamma_{nm}\b{  \hat{G}_{nm}\bullet\hat{G}_{nm}^\dagger-\f{1}{2}\left\{\hat{G}_{nm}^\dagger\hat{G}_{nm},\bullet \right\}}-
    \sum_{n=1}^{N}\lambda_{j}\sb{\hat{V}_j,\sb{\hat{V}_j,\bullet}}~~,
    \label{eq:D}
\end{equation}
where $\hat{V}_n$  are Hermitian operators, which are a linear combinations of the projection operators $\{\hat{\Pi}_j\}$. The kinetic coefficients are non-negative under Markovian dynamics and can be determined by the fixed point of the device's dynamical map $\hat{\rho}_{D}^{f.p}$, which satisfies ${\cal L}\sb{\hat{\rho}_{D}^{f.p}}=0$. For example, when the fixed point is a thermal state of inverse temperature $\beta$, the coefficients satisfy the detailed balance relation: $\gamma_{nm}=\gamma_{mn} e^{-\beta \omega_{nm}}$ \cite{dann2020open}. In other cases, such as a non-thermal bath or when the environment is composed of two thermal baths with different temperatures, the {ratio between dependent} coefficients {may be different}. {We refer to Eq. \eqref{eq:D} as the {\emph{autonomous master equation}}}. 

{Thermodynamically, the first term of $\cal{D}$ is in charge of heat transfer between the system and environment, generating transitions between the energy states of $\hat{H}_D$. Whereas the second sum is related to transfer of information, which induces dephasing in the system's energy basis. The commutation of the maps, Eq. \eqref{eq:maps_commute} implies that the  two contributions are independent \cite{dann2020open}. }

The structure of the dissipator in Eq. \eqref{eq:D} allows deriving a general expression for  $\hat{\rho}_{D}^{f.p}$. {In order to derive the general form we utilize the} orthogonality of the Lindblad jump operators $\{\hat{G}_{nm}\}${, which} implies that the first term in $\cal{D}$  can be decomposed into independent channels. Each independent channel {corresponds} to transitions between two energy states {$\ket{\psi_n}$} and {$\ket{\psi_m}$ of the device} and includes the two terms proportional to $\gamma_{nm}$ and $\gamma_{mn}$ \footnote{The transition {$\ket{\psi_m}\ra \ket{\psi_n}$} is induced by the term proportionate to $\gamma_{nm}$ and the reverse transition is associated with term proportionate to $\gamma_{mn}$.}. We denote the channel transferring populations between $\ket{\psi_n}$ and $\ket{\psi_m}$ {eigen}states of $\hat{H}_D$ as the ``$nm$-channel". 
%\tg{The Markovianity of the map, Eq. \eqref{eq:cal_Lambda_t}, implies that the kinetic coefficients are positive. In turn, this}
{The positivity of the kinetic coefficients } allows expressing the dual kinetic coefficients as $\gamma_{mn}=\gamma_{nm}e^{-\eta_{nm}}$, where $\eta_{nm}$ is an appropriate real number.
This implies that the fixed point of the $nm$-channel is of the form
\begin{equation}
    \hat{\rho}^{f.p,nm}_{D}=e^{-\bar{H}_{nm}}~~,
\end{equation}
where $\bar{H}_{nm}=\f{\eta_{nm}}{2}\b{\hat{G}_{nm}^\dagger\hat{G}_{nm}-\hat{G}_{nm}\hat{G}_{nm}^\dagger}$. Since the independent channels commute the fixed point of the entire map becomes $\hat{\rho}_D^{f.p}=Z^{-1}e^{-\bar{H}}$ with
\begin{equation}
    \bar{H}=\sum_{nm:n>m}\bar{H}_{nm}~~,
\end{equation}
and $Z$ is the partition function (cf. Appendix \ref{apsec:i.a.}).

Overall, the thermodynamic conditions determine the form of the dynamical semi-group generator up to a scaling of independent kinetic coefficients.
%\tg{Thermodynamically, the first term of $\cal{D}$ is in charge of heat transfer between the system and environment, generating transitions between the energy states of $\hat{H}_D$. Whereas the second sum is related to transfer of information, which induces dephasing in the system's energy basis. The commutation of the maps, Eq. \eqref{eq:maps_commute} infers that the  two contributions are independent \cite{dann2020open}. }

\subsection{The role of initial coherence on the open system dynamics}
\label{subsec:role_coherence}

{The presence of coherence with respect to the energy basis is related to the time-dependence of the ensuing dynamics. Specifically, sufficiently large initial coherence in the control implies coherence in the global framework, leading to transient dynamics. This initial coherence  will eventually lead to time-dependence of the primary-system in the semi-classical limit \ref{sec:semi_classical_limit}.}

Coherence with respect to the global Hamiltonian Eq. \eqref{eq:compsite_Hamiltonian} is a conserved
quantity, and therefore {may be regarded as} a resource
\cite{marvian2012symmetry,marvian2014modes,lostaglio2015quantum,streltsov2017colloquium}. This property can be understood by the following analysis. The global unitary dynamics conserves the eigenvalues (populations) of the global state $\hat{\rho}$. Hence, any function of the population is a constant of motion, such as the von-Neumann entropy ${\cal S}_{\text{VN}}=-\tr\b{\hat{\rho}\text{ln}\hat{\rho}}$. In addition, strict energy conservation and the static nature of the Hamiltonian implies that the energy entropy in the basis of $\hat{H}_D+\hat{H}_E$ is also a constant of motion ${\cal{S}}_E=-\sum_i p_i \text{ln}p_i$. From these two properties we infer the conservation of coherence, which is defined in the present study as the quantum relative entropy between the state $\hat{\rho}$ and its diagonal in the energy representation $\hat{\rho}_d$ \cite{baumgratz2014quantifying,streltsov2017colloquium}:
\begin{equation}
    {\cal D}\b{\hat{\rho}||\hat{\rho}_d}={\cal S}_E-{\cal S}_{\text{VN}}~~.
\end{equation}
In addition, an even stricter conservation condition is satisfied, namely the coherence in each mode of coherence is conserved, where different coherence modes correspond to the off-diagonal components in the energy basis which oscillate at different frequencies \cite{marvian2014modes}
%\tg{This property stems from the static nature of the Hamiltonian. In this case, 
%diagonal elements of the global state $\hat \rho$ in the energy representation are constants of motion.
%In addition, the eigenvalues of $\hat \rho$ are also conserved under unitary transformations. As a result, any measure of coherence also
 %becomes constant of motion}

{The driving force of the control mainly originates from its initial coherence. During the evolution this coherence is effectively transferred to the system, inducing a local change of energy in the system. In the local description of the primary system, this effect is manifested by a modification of the open system reduced dynamics. }
%The initial coherence in the {control} state induces transitions between the energy eigenstates of the primary system, acompanied by formation of coherence in the system.
% The initial coherence in the control with respect to $\hat{H}_C$, is approximately equal to the global coherence when $||\hat{H}_{SC}||\ll||\hat{H}_C||$. This coherence is partially transferred to the primary system, influencing the open system dynamics.
The transfer and role of coherence
is illuminated by analysing the form of the autonomous eigenoperators $\{\hat{G}_{nm}\}$. First  the general case is analyzed and  an explicit example in described in Sec. \ref{sec:JC_model}.
The autonomous eigenoperators are transition operators, $\{\ket{\psi_n}\bra{\psi_m}\}$, between the energy eigenstates of the device, $\{\ket{\psi_k}\}$. In addition, due to the interaction between the primary-system and the control ($\hat{H}_{SC}$) the device's eigenstates are generally composed of  a superposition of the bare primary-system and control eigenstates $\{\ket{\phi_i}\}$ and $\{\ket{\xi_j}\}$. This implies that a generic term of the autonomous master equation has the structure
\begin{equation}
    \hat{G}_{nm}\bullet\hat{G}_{nm}^\dagger= \ket{\psi_n}{\bra{\psi_m}\bullet \ket{\psi_m}\bra{\psi_n}=\ket{\psi_n}\b{\sum_{ij}\bra{\phi_i,\xi_j}\bullet\ket{\phi_{i'},\xi_{j'}}}}\bra{\psi_n}~~.
    \label{eq:9}
\end{equation}
{The} dynamics of the reduced primary-system is obtained by substituting  the device state instead of $\bullet$, and then tracing over the control degrees of freedom. As is observed  in Eq. \eqref{eq:9}, this procedure leads to a sum of terms which are proportionate to $\bra{\xi_j}\hat{\rho}_C\b t\ket{\xi_{j'}}$. These terms will contribute 
to the system dynamics provided the {control}
posses coherence in the control energy eigenstates for  $j\neq j'$.

\section{From autonomous description 
to an external time dependent drive - Semi-classical limit.}
\label{sec:semi_classical_limit}

When the control apparatus is sufficiently large with respect to the primary-system the autonomous framework can be replaced by a more concise semi-classical description. In this limit, the influence of the control on the primary-system is manifested by an explicit time-dependence of the composite Hamiltonian
\begin{equation}
    \hat H^{s.c}\b t={\hat{H}^{s.c}_S\b t}+\hat H_{DE}+\hat{H}_E~~.
    \label{eq:s.c_Hamil_intro}
\end{equation}

The transition between the autonomous description to the semi-classical description is termed the {\emph{semi-classical limit}}.    
The semi-classical limit is defined by the following two conditions:
\begin{enumerate}
    \item  The control state is only slightly affected by the interaction with the primary-system.
    \item The correlations between the primary-system and control are negligible, allowing to express the device's state as a separable state $\hat{\rho}_D\b t=\hat{\rho}_S\b t\otimes\hat{\rho}_C\b t$, where $\hat{\rho}_S$ and $\hat{\rho}_C$ are the primary-system and control reduced states.
    %\item The control state is effectively an eigenstate of the control terms in the interaction Hamiltonian. The primary-system control interaction can be generally expressed as of a sum over products of primary-system and control operators. $\hat{H}_{SC}=\sum_i \hat{S}_i\otimes\hat{C}_i$. In the semi-classical limit $\hat{C}_i\hat{\rho}_C\b t=\hat{\rho}_C\b t\hat{C}_i=c_i\b t\hat{\rho}_C\b t$  where $c_i\b t$ are $c$-functions.
\end{enumerate} 
The first condition is satisfied when the control is initialized in a highly excited state and the interaction with the system is negligible relative to the typical control energy scale, i.e. $||\hat{H}_{SC}||\ll||\hat{H}_C||$. In this regime, the dynamics of the control are dominated by the control Hamiltonian, indicating that $\hat{U}_D\hat{\rho}_C\hat{U}_D^\dagger\approx\hat{U}_C\hat{\rho}_C\hat{U}_C^\dagger$.
This consequence is closely related to the second condition, since if the primary-system and control are initialized in separable state, the first condition implies  that the states remain approximately separable. In addition, if we assume that the effective primary-system control coupling is of the order of the primary-system typical energy, then the magnitude of the coupling strength must be scaled accordingly, %\tg{Since $||\hat{C}_i\hat{\rho}_C||$ is very large when} 
{as} the control resides in a very energetic state. As a result, as the initial energy of the control increases the influence of the coupling term on the control state decreases, for an explicit example of this relation see Sec. \ref{sec:JC_model}.

By employing the two conditions above, we next derive the semi-classical description from the autonomous framework. 
In the autonomous description, the isolated {device} dynamics are governed by the  propagator {$\hat{U}_D\b{t,0}\in{\mathfrak{D}}$}, which is generated by the composite time-independent Hamiltonian $\hat{H}_D$, Eq. \eqref{eq:autonomous_hamil}.
By expressing the dynamics in an interaction picture relative to the bare control Hamiltonian, the autonomous propagator can be cast into the form (cf. Appendix \ref{secap:propagators})
\begin{equation}
    \hat{U}_D = \hat{U}_C\tilde{U}_D~~,
    \label{eq:relation1}
\end{equation}
where $\hat{U}_C\b{t,0}=e^{-i\hat{H}_C t/\hbar}$ is the propagator of the control and $\tilde{U}_D$ is the propagator in the interaction picture.
Written explicitly,
$\tilde{U}_D\b{t,0} = {\cal T}\exp\b{-\f{i}{\hbar}\int_0^t \tilde{H}_D\b{\tau}d\tau}\in {\mathfrak{D}}$
is generated by the effective Hamiltonian in the interaction picture relative to $\hat{H}_C$
\begin{equation}
    \tilde{H}_D\b t = \hat{H}_S+\hat{U}_C^\dagger\b{t,0} \hat{H}_{SC}\hat{U}_C\b{t,0}~~,
    \label{eq:interaction_picture_Hamil}
\end{equation}
introducing the chronological time-ordering operator ${\cal T}$ \cite{dyson1949s}. 
%Operators in the interaction picture are designated by a superscript tilde: $\tilde{X}_C =\hat{U}_C^\dagger\hat{X}_C\hat{U}_C\in{\mathfrak{C}}$.
%\tg{Combining the two semi-classical conditions we obtain the relation (cf. Appendix \ref{apsec:sc_limit})}
%\section{Semi-classical limit}
%\label{apsec:sc_limit}
%In the semi-classical limit, the  control induced dynamics  is encapsulated in terms of a $c$-function in the primary-system Hamiltonian. In the present section we derive the transition to the semi-classical description by giving a detailed derivation of the reduced dynamic of primary-system, Eq. \eqref{eq:sc_dynamics}. Relation \eqref{eq:sc_dynamics} states that the primary-system's reduced dynamics becomes:
%\begin{equation}
 %   \hat{\rho}_S\b t = \text{tr}_C\b{\hat{U}_D\hat{\rho}_D\b 0\hat{U}_D^\dagger}={\hat{U}_S^{s.c}\hat{\rho}_S\b 0\hat{U}_S^{\dagger s.c}}~~,
%\end{equation}
%where the semi-group propagator Eq. \eqref{eq:U_sc} is generated by the semi-classical Hamiltonian $\hat{H}_S^{s.c}\b t=\text{tr}_C\b{\tilde{H}_D\tilde{\rho}_C\b t}$, see Appendix \ref{apsec:sc_limit}.
%Consider the autonomous device state $\hat{\rho}_D\b t$, including both the primary-system and control.
{The corresponding device dynamics (in the interaction picture) are generated by the Liouville von-Neumann equation
\begin{equation}
    \f{d}{dt}\tilde{\rho}_D\b t = -\f{i}{\hbar}\sb{\tilde{H}_D\b t,\tilde{\rho}_D\b t}
    =-\f{i}{\hbar}\sb{\hat{H}_S+\sum_i\hat{S}\otimes\tilde{C}_i\b t,\tilde{\rho}_D\b t}~~,
    \label{apeq:eqnew}
\end{equation}
where operators in the interaction picture are designated by superscript tilde: $\tilde{X}_C=\hat{U}_C^\dagger\hat{X}_C\hat{U}_C$ (except for the Hamiltonian which is given in Eq. \eqref{eq:interaction_picture_Hamil}). In the second equality of Eq. \eqref{apeq:eqnew} we explicitly expressed the interaction Hamiltonian $\hat{H}_{SC}$ in terms of the primary-system and control operators $\{\hat{S}_i\}$ and $\{\hat{C}_i\}$.
A formal integration of the dynamical equation and  tracing over the control degrees of freedom leads to
\begin{equation}
    \hat{\rho}_S\b t = -\f{i}{\hbar}\int_0^t\text{tr}_C\b{\sb{\hat{H}_S+\sum_i\hat{S}_i\otimes\tilde{C}_i\b \tau,\tilde{\rho}_D\b{\tau}}}d\tau~~.
    \label{eq:15rho}
\end{equation}
Next, we utilize the first and second semi-classical conditions, namely: $\hat{\rho}_D\b t=\hat{\rho}_S\b t\otimes\hat{\rho}_C\b t$, with $\hat{\rho}_C\b t=\hat{U}_C\hat{\rho}_C\b 0\hat{U}_C^\dagger$, and the cyclic property of the trace to obtain 
\begin{equation}
    \hat{\rho}_S\b t = -\f{i}{\hbar}\int_0^t\sb{\hat{H}_S+\sum_i\hat{S}\otimes\text{tr}_C\b{\tilde{C}_i\b \tau\tilde{\rho}_C\b{\tau}},\hat{\rho}_S\b{\tau}}d\tau
    = -\f{i}{\hbar}\int_0^t{\sb{\hat{H}_S^{s.c}\b \tau,\hat{\rho}_S\b{\tau}}}d\tau~~.
\end{equation}}
{The differential form of this relation indicates that in the semi-classical limit the primary-system dynamics (isolated from the environment) are generated by the  Hamiltonian 
\begin{equation}
    \f{d}{dt}\hat{\rho}_S\b t = -\f{i}{\hbar}\sb{\hat{H}_S^{s.c}\b t,\hat{\rho}_S\b t}~~.
\end{equation}
where  $\hat{H}^{s.c}_S\b t=\text{tr}_C\b{\tilde{H}_D\b{t}\tilde{\rho}_C}\in {\mathfrak{S}}$ is the {\emph{semi-classical Hamiltonian}}. Note that for a highly energetic control state $\text{tr}_C\b{\tilde{C}_i\b \tau\tilde{\rho}_C\b{\tau}}$ is very large, therefore, if the primary-system-control coupling is on the scale of the system energy the coupling amplitude should be weak. } 

{Alternatively, the semi-classical propagator $\hat{U}_S^{s.c}$ satisfies the same differential equation, which leads to the formal expression} 
\begin{equation}
    \hat{\rho}_S\b t=\text{tr}_C\b{\hat{U}_D\hat{\rho}_{D}\b 0\hat{U}_D^\dagger}\cong{\hat{U}_S^{s.c}\hat{\rho}_S\b 0\hat{U}_S^{\dagger s.c}}~~,
    \label{eq:sc_dynamics}
\end{equation}
with 
 \begin{equation}
    \hat{U}^{s.c}_S\b{t,0}  = {\cal T}\exp\b{-\f{i}{\hbar}\int_0^t \hat{H}^{s.c}_S \b{\tau}d\tau}\in {\mathfrak S}~~,
    \label{eq:U_sc}
\end{equation}
where the symbol $``\cong"$ signifies an equality which is satisfied only asymptotically, in the semi-classical limit. 
Overall, the two semi-classical conditions and Eqs. \eqref{eq:relation1}, \eqref{eq:sc_dynamics} and \eqref{eq:U_sc} {imply} that the composite system dynamics are governed by 
\begin{equation}
        \hat{U}_D\cong \hat{U}_C{\otimes}\hat{U}_S^{s.c}~~.
\label{eq:sc_limit_prop}
\end{equation}
This relation serves as a key identity in the analysis of the master equation in the semi-classical regime.

\section{Driven open quantum system dynamics (semi-classical regime)}
\label{sec:open_driven_system}

Combining the autonomous dynamical symmetry Eq. \eqref{eq:maps_commute}, with the semi-classical procedure introduced in Sec. \ref{sec:semi_classical_limit}, we next derive the reduced dynamics of a driven open quantum system.
%The form of the semi-classical master equation \eqref{eq:D_prime} can be related to time-translation symmetry of the evolution.
Evaluation of the semi-classical limit of Eq. \eqref{eq:maps_commute} leads to 
the commutation relation of the open and isolated semi-classical dynamical maps (see Appendix \ref{apsec:time-translation} for a detailed derivation)  
\begin{equation}
    \Lambda^{s.c}\circ{\cal U}_{S}^{s.c}={\cal U}_{S}^{s.c}\circ\Lambda^{s.c}~~,
\label{eq:s.c_commut}
\end{equation}
where $\Lambda^{s.c}\sb{\bullet}\b t=\text{tr}_{E}\b{U_{SE}^{s.c}\b t\bullet U_{SE}^{s.c\dagger}\b t}$.
Here, $\hat U_{SE}^{s.c}$ is the semi-classical combined time-evolution operator, which acts on the Hilbert space of the primary-system and environment. This operator satisfies the Schr\"odinger equation with respect to the semi-classical composite Hamiltonian $\hat{H}^{s.c}\b t=\text{tr}_C\b{\hat{H}\hat{\rho}_C \b t}$.

The commutation relation of the semi-classical dynamical maps Eq. \eqref{eq:s.c_commut} sets strict restrictions on the maps structure. Assuming the spectrum of ${\cal U}_S^{s.c}$ is non-degenerate, the two maps share the same non-invariant eigenoperators, sharing an analogous structure as in the autonomous case. Despite of the similarity there is an important difference between the autonomous and semi-classical maps. Namely, the autonomous maps self-commute at different times, while the semi-classical may not.    This implies that the semi-classical dynamical generators do not necessarily commute with the maps. As a result, the structure of the semi-classical generator, ${\cal L}^{s.c}_S$, which is defined by the relation 
\begin{equation}
    \f{d}{dt}\Lambda^{s.c}\b{t}={\cal{L}}^{s.c}_S\b{t}\circ\Lambda^{s.c}\b{t}~~,
    \label{eq:s.c_diff}
\end{equation}
cannot be deduced in a straight forward manner from the structure of $\Lambda^{s.c}\b t$. Note, that to obtain ${\cal L}^{s.c}_S$ from Eq. \eqref{eq:s.c_diff} requires a time-ordering procedure when $\Lambda^{s.c}\b t $ does not commute with itself at different times \cite{dyson1949s}.
However, the issue can be resolved by employing a time-dependent basis to represent the isolated and open system maps in Liouville space in the Heisenberg picture.
By choosing a suitable time-dependent operators basis the free dynamical generator is represented by a Liouville space superoperator which commutes with it self at different times. In this representation the open system map commutes with the generator, leading to the operatorial structure of the dynamical generator.

We begin the derivation by studying the eigenvalue solutions to the Heisenberg equation   
\begin{equation}
    \f{d}{dt}\hat{X}_S^H\b t \equiv{\cal G}_S^{s.c}\b t\b{\hat{X}_S\b t}= \hat{U}_S^{s.c\dagger}\b{t}\b{\f{i}{\hbar}\sb{\hat{H}_S^{s.c}\b t,\hat{X}_S\b t}+\pd{\hat{X}_S\b t}{t}}\hat{U}_S^{s.c}\b t~~. 
    \label{eq:hies}
\end{equation}
The eigneoperators (of the Hiesenberg representation) $\{\hat{P}_\alpha\b t\}$ are time-dependent operators which satisfy  
the eigenvalue equation with respect to the free dynamical generator 
\begin{equation}
    \f{d}{dt}\hat{P}^H_k\b t = i \lam_k \b t\hat{P}^H_k \b t~~.
    \label{eq:eigoper}
\end{equation}
where the superscript $H$ designates operators in the Heisenberg picture, and $\lam_k$ are real scalars. The set constitutes a complete basis, they therefore can be utilized to solve the dynamics of any system operator. For a general drive it may be difficult to obtain a closed analytical form for $\{\hat{P}_k\b t\}$, nevertheless their existence is guaranteed for an arbitrary drive \cite{dann2021inertial}. In Sec. \ref{sec:dete_eig} we propose a numerical method, employing a Fourier transform of Eq. \eqref{eq:hies} by which to evaluate the eigenoperators. 

In analogy with the autonomous case, the set of eigenoperators $\{\hat{P}_k\b t\}$ are composed on non-invariant operators $\{\hat{F}_k\}$, for which $\lam_k\neq 0$, and invariant eigenoperators $\{\hat{W}_j\}$ with vanishing eigenvalues. Therefore, if we choose the set of eigenoperators as the operator basis $\{\hat{F}_1\b t,\dots,\hat{F}_{N\b{N-1}}\b t,\hat{W}_1\b t,\dots,\hat{W}_N\b t\}$, the semi-classical propagator in the Heisenberg representation, ${\cal U}_S^{s.c\ddagger}\b t\sb{\bullet}={\cal{T}}e^{-i\int_0^t{\cal{G}}_S^{s.c}\b{s}\sb{\bullet}ds}$, is of a diagonal form 
\begin{equation}
    \hat{P}_k^H\b t={\cal U}_S^{s.c\ddagger}\b t\sb{\hat{P}_k\b 0} =\exp \b{i\int_0^t\, \lam_k\b{s}ds}\hat{P}_k\b 0~~.
    \label{eq:25}
\end{equation}
Here, the Heisenberg and Schr\"odinger operators coincide at initial time, i.e., $\hat{P}_k^H\b 0=\hat{P}_k\b 0$, and the superscript $\ddagger$ designates superoperators in the Heisenberg picture. The relation implies that $\{\hat{P}_k\b 0\}$, constitute eigenoperators of the free map. This property along with the dynamical symmetry relation, Eq. \eqref{eq:s.c_commut}, enables deducing the exact form of the semi-classical master equation.

We follow the same deduction as in the autonomous framework Sec. \ref{sec:time_independent}. When the eigenvalues of the non-invariant operators are non-degenerate, the symmetry relation infers that they constitute eigenoperators of the semi-classical open system map $\Lambda^{s.c\ddagger}\b t$
\begin{equation}
    \hat{F}_k^H\b t={ \Lambda}^{s.c\ddagger}\b t\sb{\hat{F}_k\b 0} =\eta_k\b t\hat{F}_k\b 0~~.
\end{equation}
where $\eta_k$ are complex numbers. 

Finally, the lack of time-dependence of the eigenoperators in this representation allows obtaining the form of the semi-classical dynamical generator from  Eq. \eqref{eq:s.c_diff}.  The relation 
$   {\cal L}_S^{s.c}\b t =\b{\f{d}{dt}\Lambda^{s.c\ddagger}\b t}\b{\Lambda^{s.c
\ddagger}\b t}^{-1}$ infers that the generator is diagonal in the non-invariant eigenoperators, ${\cal L}_S^{s.c\ddagger}\sb{\hat{F}_k\b 0} = \dot{\eta}_k \eta_k^{-1}\hat{F}_k\b 0$, and maps the invariant subspace to itself: ${\cal L}_S^{s.c}\sb{\hat{W}_i\b 0} = \sum_j a_{ij}\b t\hat{W}_j\b 0$, where $a_{ij}$ are time-dependent coefficients.
Under the four thermodynamic postulates, these relations impose the following structure for the semi-classical dissipator \cite{dann2020open}
\begin{equation}
    {\cal{D}}^{s.c}\sb{\bullet} = \sum_{k}\gamma_k\b t\b{  \hat{F}_{k}\b 0\bullet\hat{F}_{k}^\dagger\b 0-\f{1}{2}\left\{\hat{F}_{k}^\dagger\b 0\hat{F}_{k}\b 0,\bullet \right\}}+
    \sum_{ij}\chi_{ij}\b t\hat{W}_i\b 0\bullet\hat{W}_j\b 0~~.
    \label{eq:D_prime}
\end{equation}

The main feature of the semi-classical master equation is the connection between the unitary and dissipative parts. The Lindblad jump operators constitute eigenoperators of the driven (isolated) system and therefore depend on the drive.  This relation contrasts with the standard description in  which the environment and drive generate independent terms in the dynamical generator. Equation \eqref{eq:D_prime} demonstrates that the underlying quantum nature of the drive leads an inevitable dependency between the external affects. 

The structure of the master equation is similar to the autonomous case Eqs. \eqref{eq:L_decompos} and \eqref{eq:D}: The semi-classical dissipator is composed of two terms. The first is related to energy transitions between the eigenstates of $\hat{U}^{s.c}_S$, which are induced by the jump operators $\{\hat{F}_k\}$. While the second term leads to dephasing in this eigenstates' basis and also includes a unitary correction to the effective energy levels due to the presence of the environment, the so-called Lamb-shift term. 
Under Markovian dynamics the $\{\gamma_k\b t\}$ are necessarily positive. A similar structure appears under non-Markovian dynamics, while the coefficients may obtain negative values \cite{dann2021non}.

%{In the Schr\"odinger representation Eq. \eqref{eq:D_prime} can be decomposed to time-dependent unitary and dissipative terms. These two terms are connected, as the Lindblad jump operators $\hat F_k\b t$  are also eigenoperators of the semi-classical propagator $\hat{U}_{S}^{s.c}$. Thus, the free propagation determines the dissipative part.

Interestingly, in the semi-classical framework the fixed point of the map is replaced by a time-dependent instantaneous attractor $\hat{\rho}_S^{i.a}\b t$. This operator satisfies the relation ${\cal{L}}^{s.c}\b{t}\sb{\hat{\rho}_S^{i.a}\b t}=0$ instantaneously. Typically, the instantaneous attractor may not commute with the semi-classical Hamiltonian, which implies that energy and coherence in the instantaneous eigenbasis of $\hat{H}_S^{s.c}\b t$ mix \cite{dann2018time}.

%\tb{connect between the sections}
%The entropy production rate is identified with $\sigma\b{\tilde{\rho}_S\b t}=-\f{d}{dt}S\b{\tilde{\rho}_S\b t||\hat{\rho}_S^{i.a}\b t}$, where $S\b{\hat{\rho}||\hat{\sigma}}\equiv\text{tr}\b{\hat{\rho}\text{ln}\hat{\rho}-\hat{\rho}\text{ln}\hat{\sigma}}$ is the quantum relative entropy. In turn, the positivity of $\sigma$ under a CPTP map leads to the quantum version of the $II$ law \cite{spohn1978irreversible,alicki1979quantum}.

Overall, the present construction shows that the restrictions on the structure of the master equation, in the semi-classical regime, is a manifestation of the strict energy conservation between the {device} and the environment.

\section{General form of the instantaneous attractor}

The identification of the Lindblad jump operators as the eigenoperators of the free dynamics allows obtaining a general form for the instantaneous attractor.

We first express the dissipator the interaction picture relative to the free dynamics and omit the time-dependence from the  kinetic coefficients and eigenoperators throughout this section, i.e, $\hat{F}_k\equiv\b 0\equiv \hat{F}_k$, $\hat{W}_j\equiv\b 0\equiv \hat{F}_j$  
\begin{equation}
    \widetilde{\cal{D}}^{s.c}\sb{\bullet} = \sum_{k}\gamma_k\b{  \hat{F}_{k}\bullet\hat{F}_{k}^\dagger+\f{1}{2}\left\{\hat{F}_{k}^\dagger\hat{F}_{k},\bullet \right\}}+
    \sum_{ij}\chi_{j}\hat{W}_i\bullet\hat{W}_j~~.
    \label{eq:new_eq}
\end{equation}
{For a finite level system, the non-invariant eigenoperators $\{\hat{F}_k\}$ are effective transition operators in the primary system-control dressed basis,
they therefore satisfy the following properties: $\hat{F}_k^2=\hat{0}$ and $\tr\b{\hat{F}_k\hat{F}_l}=\delta_{kl}$. These
properties imply that the first term of Eq. \eqref{eq:new_eq} is composed of independent channels, where} the $k$'th channel reads 
\begin{equation}
    \widetilde{\cal{D}}_S^{s.c\b{k}}\sb{\bullet} = \Gamma_k\b{  \hat{F}_{k}\bullet\hat{F}_{k}^\dagger-\f{1}{2}\left\{\hat{F}_{k}^\dagger\hat{F}_{k},\bullet \right\}}+\Gamma_{-k}\b{  \hat{F}_{k}^\dagger\bullet\hat{F}_{k}-\f{1}{2}\left\{\hat{F}_{k}\hat{F}_{k}^\dagger,\bullet \right\}}~~.
    \label{eq:eq28}
\end{equation}
Physically, such a channel corresponds to a pathway for transfer of probability between two distinct states of the primary-system. These channels can be thought as independent connections between the primary-system and environment. 
In addition, the invariant operators $\{\hat{W}_j\}$ can be expressed in terms of the projection operators, suggesting that the second term of Eq. \eqref{eq:new_eq}  vanishes for  $\bullet =\hat{F}_k^\dagger\hat{F}_k$ or $\hat{F}_k\hat{F}_k^\dagger$. 
These {considerations} motivate defining an effective Hamiltonian 
\begin{equation}
    \bar{H}_{S} =\sum_k \f{\delta_k}{2}\b{\hat{F}_k^\dagger \hat{F}_k-\hat{F}_k\hat{F}_k^\dagger }~~,
    \label{eq:eff_Ham}
\end{equation}
with undetermined real parameters $\delta_k$.
Such an operator is characterized by the simple commutation relation
\begin{equation}
    \sb{\bar{H}_S,\hat{F}_k}=- \delta_k \hat{F}_k~~.
    \label{eq:comutation_eff}
\end{equation}
In analogy with the autonomous case, these commutation relations suggest that the instantaneous attractor is of the form 
\begin{equation}
 \widetilde{\rho}_S^{i.a}\b t=Z^{-1} \exp\b{-\bar{H}_S\b t}~~,
 \label{eq:instanteneous_attractor}
\end{equation}
where $Z$ is the corresponding partition function. Substituting Eq. \eqref{eq:instanteneous_attractor} into Eq. \eqref{eq:new_eq}, determines the parameters $\delta_k=\text{ln}\b{\Gamma_{k}/\Gamma_{-k}}$ (see Appendix \ref{apsec:i.a.} for a detailed derivation).

\section{Determining the eigenoperators}
\label{sec:dete_eig}

The Lindblad jump operators of the master equation constitute eigenoperators of the dynamical generator ${\cal{G}}_S^{s.c}$, Eq. \eqref{eq:hies}. 
Obtaining the diagonalizing basis is not straightforward as the generator ${\cal G}_S^{s.c}$ is generally time-dependent \footnote{A naive diagonalization of the differential equation $\f{d\v{V}}{dt}={\cal{G}}_S\b t \v V\b t$, with the instantaneous diagonalizing matrices ${\cal P}\b t$ leads to  $\f{d}{dt}\b{{\cal P}^{-1} \v{V}}=\sb{{\cal{D}}_S +\f{d{\cal P}^{-1}}{dt}{\cal P}}{\cal P}^{-1}\v V\b t$, where ${\cal D}_S={\cal P}^{-1}{\cal G}_S{\cal P}$ is the a diagonal matrix. When the generator does not commute with itself at different times, the additional term $\f{d{\cal P}^{-1}}{dt}{\cal P}$ may be non-diagonal and the set of differential equations remain coupled.}.  Nevertheless, this complication can be bypassed by transforming to the frequency domain. Performing a Fourier transform of Eq. \eqref{eq:hies}, the eigenvalue equation for $\hat{F}_k^H\b t$   can expressed as
\begin{equation}
    \b{\sb{\hat{H}_S\b{\omega},\bullet}+\omega}\hat{F}_{k}^H\b{\omega}=-\lam_k\b{\omega}\hat{F}_{k}^H\b{\omega}~~,
    \label{eq:eig_value_freq}
\end{equation}
where $\hat{F}_k\b{\omega}$ is an eigenoperator in the frequency domain. Finally, we express Eq. \eqref{eq:eig_value_freq} in a matrix vector form by performing a vec-ing procedure $\hat{F}_k^H\b \omega\ra\v f_k\b \omega$ (cf. Appendix \ref{apsec:unitity_of_U}) (here any choice of operator basis will also do). This leads to the following eigenvalue equation  
\begin{equation}
    \b{\hat{I}\otimes\hat{H}_S-\hat{H}^{T}_S\otimes\hat{I}_S-\omega\hat{I}_S\otimes\hat{I}_S}\v f_{k}\b{\omega}=-\lam_k\b{\omega}\v f_{k}\b{\omega}~~,
\end{equation}
where $\v{f}_k$ are vectors in Liouville space. The solution to the eigenvalue problem gives the eigenvectors in Liouville space $\v{f}_k\b t$ and the corresponding  eigenoperators $\hat{F}_k^H\b t$. For a general system Hamiltonian the diagonalization can be performed by numerical diagonalization methods to obtain the eigenoperators within the desired accuracy.

\section{Comparison with other master equations}
\label{sec:compare}

The master equation for the driven system ($\mathfrak{S}$), Eq. \eqref{eq:D_prime}, has a strict structure which emerges solely from the thermodynamic postulates. This structure is consistent with a number of master equations, derived by conducting a microscopic derivation. 
A prime example is  the non-adiabatic master equation (NAME) \cite{dann2018time}, which governs the dynamics of a driven system in contact with a Markovian bath. The procedure leading to the NAME has a similar starting point of a combined system-environment evolution, Eq. \eqref{eq:s.c_Hamil_intro}, and utilizes weak coupling-Markov-secular approximation to obtain the GKLS form. In contrast to the general structure of Eq. \eqref{eq:D_prime}, the microscopic derivation leads to an explicit form for the kinetic coefficients of the NAME. These are determined from the Fourier transforms of the environment correlation functions with a modified frequency compared to bare Bohr frequencies of the system. However, the price,
of the complete formal derivation is that the assumptions employed limit the regime of validity.

For periodic driving one can utilize the Floquet analysis to derive a consistent master equation  \cite{alicki2012periodically,szczygielski2013markovian,nathan2020universal}. This master equation is connected to the NAME, which constitutes a generalization of the Floquet master equation, see Sec. \ref{sec:JC_open} for example.

In the slow driving limit, the obtained structure corresponds to the adiabatic master equation \cite{albash2012quantum}. In this case, the free evolution is given by the adiabatic propagator $\hat{U}_S^{adi}\b t=e^{-i\hat{H}_S^{s.c}\b t t/\hbar}$, and the eigenoperators become the transitions operators of the ``instantaneous" semi-classical Hamiltonian $\hat{H}_S^{s.c}\b t$. In this driving regime and for a thermal bath with inverse temperature $\beta$, the instantaneous attractor is a Gibbs state with respect to the instantaneous Hamiltonian $\hat{\rho}_S^{i.a}\b t=Z^{-1}\b te^{-\beta\hat{H}_S^{s.c}\b t}$, where $Z\b t$ is the partition function.

An alternative approach to obtain the dynamical equations assumes the driven master equation can be approximated using a time-dependent Hamiltonian and a static dissipative part. Due to its relative simplicity, this approximation has gained popularity in many disciplines, such as, quantum optics, quantum information and spectroscopy. For example, the well-known Bloch equation has this structure \cite{geva1995relaxation}. However, there is only a limited regime where such a description complies with the thermodynamiclly consistent structure of Eq. \eqref{eq:D_prime}.
%This approach is valid in a limited regime, for example, in the singular coupling limit  \cite{breuer2002theory}, where  the structure of the master equation reduces to a double commutator in the interaction Hamiltonian \cite{gorini1976n}.
Nevertheless, in the case of a singular bath, the dissipative and unitary part may be independent. In this scenario transfer energy can occur with no entropy cost, which is only consistent with thermodynamics for a bath of infinite temperature. 
%In addition, the static approach to the master equation can be obtained in the high frequency and low intensity limit of the Floquet master equation with an additional averaging over the fastest timescale. 

\section{Example: Jaynes-Cummings model}
\label{sec:JC_model}

The Jaynes-Cummings (JC) model serves as a basic toy model in quantum optics and solid state physics. It allows a full quantum treatment of a quantum system interacting with an external field \cite{jaynes1963comparison,shore1993jaynes}. 
It describes a qubit interacting with
a single mode of a bosonic field. Since the model can be solved in closed form, it constitutes a natural example to demonstrate the transition between the autonomous and semi-classical descriptions. We identify the primary-system with the qubit while the control subsystem is represented by the bosonic mode, see Table \ref{table1} for a summary of the analogies between the notations of the general case and the example. In the appropriate limit the autonomous quantum dynamics converge to the semi-classical dynamics, characterized by  Rabi oscillations. 
There are many other possibilities to construct a control system which posses the semi-classical properties. For example, many independent synchronized units will obey the property, with the appropriate interaction terms \cite{phillips2001storage}.

Within the rotating wave approximation the JC Hamiltonian is given by \cite{shore1993jaynes}
\begin{equation}
    \hat{H}_D^{JC} = \hbar\omega_{c}\b{\hat{a}^{\dagger}\hat{a}+\f 12}+\f{\hbar\omega_{eg}}2\hat \sigma_{z}+\hbar g\b{\hat \sigma_{-}\hat a^{\dagger}+\hat \sigma_{+}\hat a}~~,
    \label{eq:JC_hamil}
\end{equation}
where $\hat{a}$ and $\hat{a}^\dagger$ are the creation anhillation operators of the bosonic mode, $\hat{\sigma}_i$, $i=x,y,z$ are the Pauli operators of the qubit, $g$ is the coupling strength and $\omega_c$ and $\omega_{eg}$ are the oscillator and qubit internal frequencies. The eigenstates of the bare qubit and bosonic mode Hamiltonians are denoted by $\ket{g/e}$ and $\ket{n}$, correspondingly.

\begin{center}
 \begin{tabular}{|c c |} 
 \hline
 General derivation & Example  \\ [0.5ex] 
 \hline\hline
 $\hat{H}_S$ & $\hbar \omega_{eg}\hat{\sigma}_z/{2}$  \\
 $\hat{H}_C$ & $\hbar \omega_c\b{\hat{a}^\dagger\hat a+\f{1}{2}}$  \\
 $\hat{H}_{SC}$ & $\hbar g\b{\hat \sigma_{-}\hat a^{\dagger}+\hat \sigma_{+}\hat a}$   \\
 $\hat{F}_k$ &  $ \hat{F}_\pm$ \\ 
  $\hat{W}_k$ &  $ {\hat{W}}$ \\
   $\hat{C}_k$ & $\{\ket{n}\bra{m}\}$   \\
   [1ex] 
 \hline
\end{tabular}
\label{table1}
\end{center}

\subsection{Semi-classical limit for an isolated system}
\label{sec:JC_isolated}

The semi-classical limit is obtained by initializing the control subsystem in a coherent state $\hat{\rho}_C\b 0=\ket{\alpha}\bra{\alpha}$ with a large average photon number $\mean{n}=\bra{\alpha}\hat{a}^\dagger\hat{a}\ket{\alpha}=|\alpha|^2\gg 1$. In addition, the effective interaction term $g\sqrt{n}$ is kept constant and weak relative to energy scale of the system $\omega_{eg}$. 

The semi-classical limit is
demonstrated  by two procedures: First {we show that the classical conditions, cf. Sec. \ref{sec:semi_classical_limit}, lead} to the semi-classical propagator $\hat{U}_S^{s.c}$, Eq. \eqref{eq:sc_limit_prop}, in terms of semi-classical Hamiltonian $\hat{H}_S^{s.c}\b t$. In the second procedure we directly derive $\hat{U}_S^{s.c}$ from the autonomous propagator $\hat{U}_D$, Eq. \eqref{eq:relation1}, in the semi-classical limit.

The first semi-classical condition, Sec. \ref{sec:semi_classical_limit}, assumes the control remains unaffected by the interaction with the primary-system. This is a consequence  of the requirement that the coupling with the primary-system remains moderate. To satisfy this restriction, $g$ must scale accordingly $g\propto 1/|\alpha|$ and becomes very small in the semi-classical limit. Therefore, when comparing the energy scale of the bare control Hamiltonian and the coupling term we find that $||\hat{H}_{SC}||\ll||\hat{H}_C||$. As a result, the contribution of the interaction term to the dynamics of the control state is negligible (does not modify $\hat{U}_D$ relative to $\hat{U}_C$ substantially).
Moreover, if we assume an initial separable state $\hat{\rho}_D\b 0=\hat{\rho}_S\b 0\otimes \ket{\alpha}\bra{\alpha}$, the first condition implies that the state remains approximately separable. Thus, the second condition is obtained as a consequence of the first condition. 

The general procedure leads to the semi-classical propagator $\hat{U}_S^{s.c}$ which is determined by its generator, the semi-classical Hamiltonian $\hat{H}_S^{s.c}\b t=\text{tr}_C\b{\tilde{H}_D\b t\tilde{\rho}_C\b t}$.
 In terms of the JC model parameters the semi-classical Hamiltonian becomes
\begin{equation}
        \hat{H}_{S}^{JC,s.c}\b t = \f{\hbar\omega_{eg}}2\hat \sigma_{z}+\hbar g\b{\hat \sigma_{-}\alpha^*e^{i\omega_c t}+\hat \sigma_{+}\alpha e^{-i\omega_c t}}~~,
    \label{eq:JC_hamil}
\end{equation}
which is the Rabi model Hamiltonian. The propagator $\hat{U}_S^{JC,s.c}$ {is obtained by a suitable change of basis}, leading to  (see Appendix \ref{apsec:comparison} for more details)
\begin{equation}
    \hat{U}_S^{JC,s.c}\b{t,0}=
    \left(\begin{array}{cc}
\text{c}_{\mean{n}}+\frac{i\Delta\text{s}_{\mean{n}}}{\Omega_{{\mean{n}}}} & -i\f{2 g\sqrt{{\mean{n}}}}{\Omega_{{\mean{n}}}}\text{s}_n\\
-i\f{2 g\sqrt{{\mean{n}}}}{\Omega_{{\mean{n}}}}\text{s}_n & \text{c}_{\mean{n}}-\frac{i\Delta\text{s}_{\mean{n}}}{\Omega_{{\mean{n}}}}
\end{array}\right)~~,
\label{eq:U_S^JC,s.c}
\end{equation}
 with $\text{c}_{\mean{n}}=\cos\left(\frac{\Omega_{{\mean{n}}}t}{2}\right)$, $\text{s}_{\mean{n}}=\sin\left(\frac{\Omega_{{\mean{n}}}t}{2}\right)$ and $\Omega_{\mean{n}}=\sqrt{\Delta^{2}+4g^{2}{\mean{n}}}$.
 The matrix in Eq. \eqref{eq:U_S^JC,s.c} is expressed in the basis $\{\ket{g},\ket{e}\}$.

%Finally, the third condition is obtained from the properties of a coherent state: $\hat{a}\ket{\alpha}=\alpha\ket{\alpha}$ and $\bra{\alpha}\hat{a}^\dagger\hat{a}\ket{a}=-1+|\alpha|^2\approx|\alpha|^2$, which suggests that $\hat{a}^\dagger\ket{\alpha}\approx\alpha^*\ket{\alpha}$. These relations imply that $\hat{H}_{SC} = \hbar g\b{\hat \sigma_{-}\hat a^{\dagger}+\hat \sigma_{+}\hat a}$ satisfies the required property: the action of the interaction Hamiltonian amounts to a multiplication by a $c$-function.

A second approach to demonstrate the convergence to the semi-classical description utilizes the explicit solution
of the JC propagator.  This enables solving for the autonomous propagator $\hat{U}_D^{JC}$ (generated by $\hat{H}_D^{JC}$) and  taking the  semi-classical limit. 
The solution is obtained by splitting the JC Hamilonian, Eq. \eqref{eq:JC_hamil}, into block diagonal matrices operating on subspaces $\{\ket{g,n},\ket{e,n-1}\}$
\begin{equation}
    \hat{H}_{D}^{\b n}=n\hbar\omega_{c}\hat I^{\b n}-\f{\hbar\Delta}2\hat \sigma_{z}^{\b n}+\sqrt{n}\hbar g\hat \sigma_{x}^{\b n}~~,
    \label{eq:H_S^n}
\end{equation}
with a detuning $\Delta=\omega_{eg}-\omega_{c}$ and Pauli operators $\hat{\sigma}_i^{\b{n}}$, operating on the effective two-dimensional $n$'th subspace.
This decomposition infers that the solution for the autonomous propagator is given by a sum over the independent propagators $\hat{U}_D^{JC}=\sum_n \hat{U}_D^{\b n}$, where $\hat{U}_D^{\b{n}}=e^{-i\hat{H}^{\b{n}}_Dt/\hbar}$.

\begin{figure}
\centering
\includegraphics[width=7.cm]{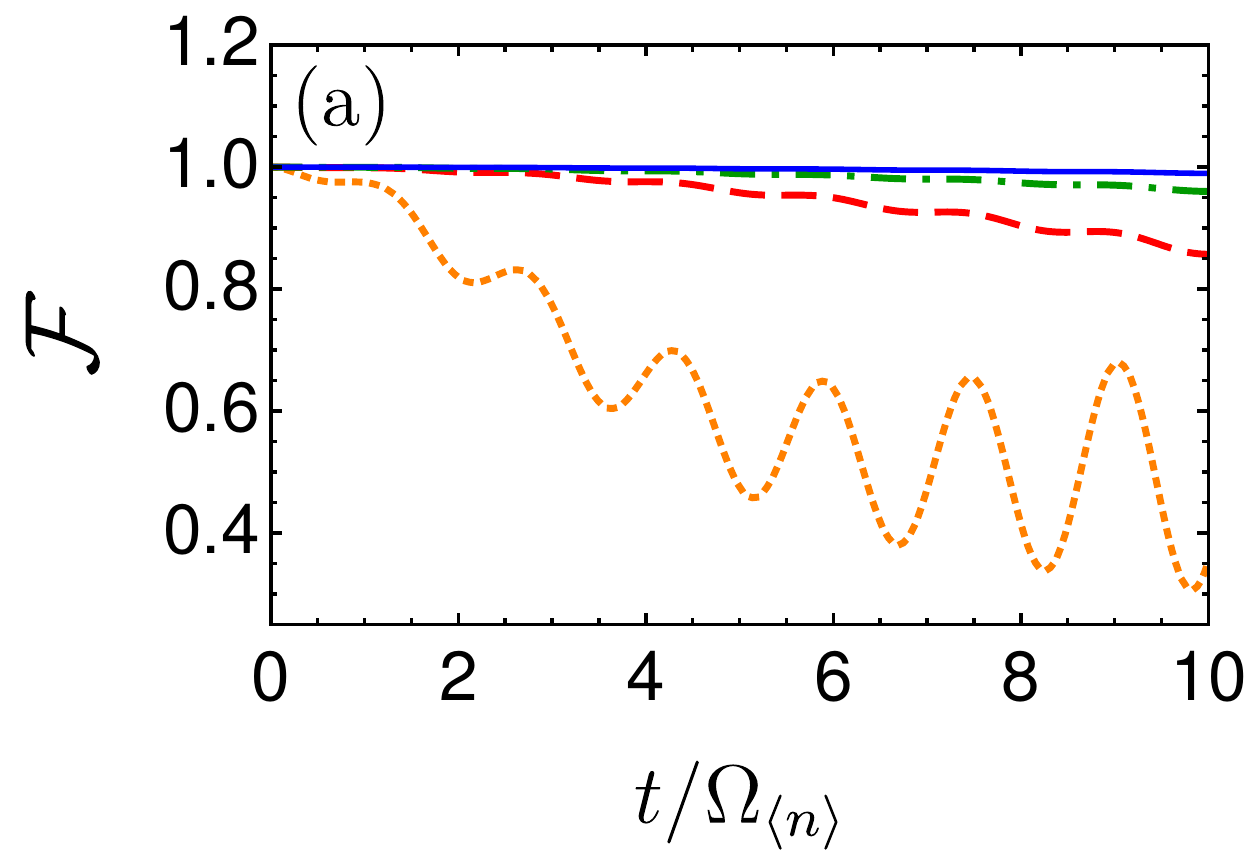}\includegraphics[width=7.cm]{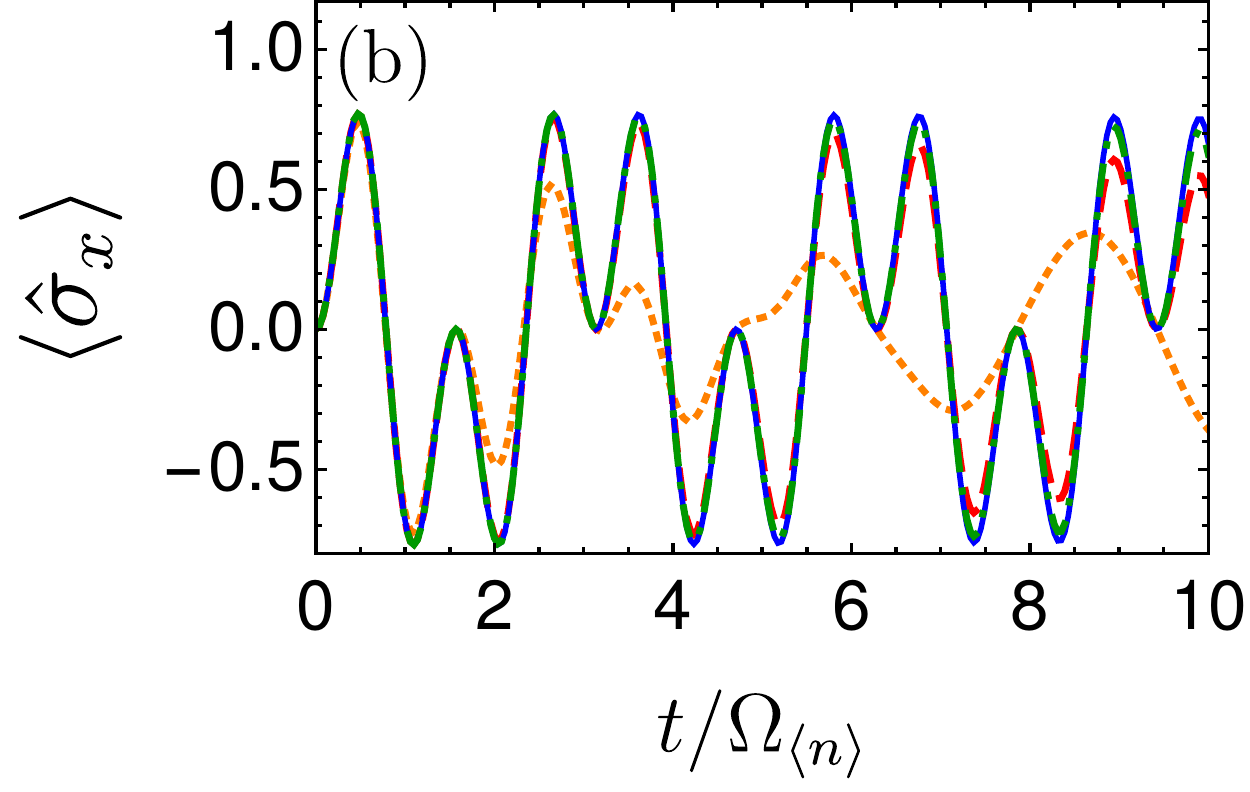}
\caption{{Convergence to the classical limit for increasing $\alpha$. Panel (a) presents the Uhlmann fidelity $\cal F$ between the solution of the Jaynes-Cumming model (autonomous description Eq. \eqref{eq:JC_hamil}) and semi-classical solution (solution of Eq. \eqref{eq:JC_hamil}) as a function of normalized time for different initial coherent states $\hat{\rho}_C \b 0 = \ket{\alpha}$: Orange dotted - $\alpha=5$; Red dashed - $\alpha = 25$; Green dashed dotted - $\alpha = 50$, Blue continuous - $\alpha=100$. The lowest value of $\alpha$ sticks out, while larger coherent states are close together and deviate only at longer times. Panel (b) exhibits the expectation value of the Pauli matrix $\hat{\sigma}_x$ as a function of time with the same color notation, similar results are obtained for the other Pauli matrices. Time is normalized by the generalized Rabi frequency of the semi-classical solution $\Omega_{\mean{n}}$. The chosen model parameters are: $\Omega_{\mean{n}}=2$, $\Delta = 0$ and $\omega_c =1$. Note that similar results are obtained in the non-resonant case, when $\Delta \neq 0$.  }
}
\label{fig:JC_rabi_comp}
\end{figure}

In the semi-classical limit the boson field is initialized in a highly energetic coherent state, $\mean{n}\gg 1$, while keeping the coupling to the qubit moderate, $g\sqrt{\mean{n}}=\text{const}<\omega_{eg}$. In this regime, the fluctuations around the mean value become negligible and the sum over propagators converges to $\hat{U}_D^{JC} \simeq \hat{U}_D^{\b{\mean{n}}}$, see Appendix \ref{apsec:convergence}. This propagator coincides with the semi-classical propagator $\hat{U}_S^{JC,s.c}$, Eq. \eqref{eq:U_S^JC,s.c}, as obtained from the first approach. {The convergence to the semi-classical limit is shown in Fig. \ref{fig:JC_rabi_comp}, presenting the Uhlmann fidelity between the autonomous and semi-classical solutions (Panel (a)) and $\mean{\hat{\sigma}_x}$ (Panel (b)) as a function of time for different initial control coherent states $\ket{\alpha}$. With the increase of $\alpha$, the Janyes-Cummings solution (autonomous) converges to Rabi oscillations result (semi-classical), manifested by a fidelity close to unity in the chosen time interval.}

%In turn, the  propagator in the interaction picture, Eq. \eqref{eq:relation1}, becomes
%\begin{equation}
 %   \tilde{U}_S = \hat{U}_C^\dagger\hat{U}_S^{\b{\mean{n}}}~~.
%\end{equation}
%In accordance with Eq. \eqref{eq:sc_limit_prop}, we recognize the semi-classical propagator as $\hat{U}_S^{s.c}=\hat{U}_S^{\b{\mean{n}}}$.

The convergence to the semi-classical limit can be studied by analysing the coherence properties for ``short" times ($gt<10$), which were first studied by Cummings \cite{cummings1965stimulated} and further generalized by Eberly et al. \cite{eberly1980periodic}. In this temporal regime, the typical oscillations are terminated by a Gaussian envelope $\exp \b{-\phi\b t}$ \cite{eberly1980periodic},
with $ \phi\b t = {\f{2\mean{n}g^2}{\Delta^2+4\mean{n}g^2}} \b{gt}^2$. Since 
$g\sqrt{\mean{n}}$ remains constant in the semi-classical limit, the loss of coherence is suppressed by the magnitude of the coherent state:
$\phi\b t\propto {|\alpha|^{-2}}$.
In  Appendix \ref{apsec:convergence} we present an alternative derivation of the convergence to the semi-classical limit,  relying on the asymptotic behaviour of the Touchard polynomials. {The eventual deviance between the autonomous and semi-classical results can be seen in Fig. \ref{fig:JC_rabi_comp} Panel (a), where the fidelity drops from unity at times for which $t>|\alpha|^{-2}$.}

To summarize, the studied example demonstrates the validity of Eqs. \eqref{eq:sc_dynamics}  and \eqref{eq:U_sc}, which determine the semi-classical dynamics. It shows that the semi-classical description can be obtained solely in terms of the semi-classical Hamiltonian $\hat{H}_S^{s.c}\b t$, by relying on the general procedure presented in Sec. \ref{sec:semi_classical_limit}. This approach serves as a simple straight forward method to obtain the semi-classical description. The resulting dynamics coincides with the result obtained by first completely solving the autonomous dynamics and then taking the semi-classical limit.
%This procedure gives a straight forward method to determine the propagator $\hat{U}_S^{s.c}$ and the primary-system dynamics $\hat{\rho}_S\b t$. 

\subsection{Eigenoperators in the semi-classical limit}

 The eigenoperators in the semi-classical limit can be derived by introducing a set of time-dependent operators $\v{v}\b t \equiv  \{\hat{\sigma}_-e^{i\omega_c t},\hat\sigma_+e^{-i\omega_c t},\hat{\sigma}_z/\sqrt 2\}$. Along with the identity operator, this set of operators constitutes a complete basis for the qubit operator vector space, the so-called Liouville space. Utilizing the Heisenberg equation of motion the dynamics of the basis operators can be expressed in a matrix-vector form: 
 $\f{d}{dt} \v{v}^{H}\b t=M \v{v}^H\b t$, where 
 \begin{equation}
     M=i\sb{\begin{array}{ccc}
-\Delta & 0 & \sqrt{2} g\alpha\\
0 & \Delta & - \sqrt{2}g\alpha^{*}\\
\sqrt{2}g\alpha^{*} & -\sqrt{2}g\alpha & 0
\end{array}}~~.
 \end{equation}
 Diagonalization of $M$ leads to three uncoupled equations of motion, governing the dynamics of the eigenoperators of $\hat{U}_S^{JC,s.c}$. The transition operators (eigenoperators with non-vanishing eigenvalues) in the Schr\"odinger picture  are given by
 \begin{equation}
 \hat{F}_{\pm}\b t=N_{\pm }\b{\b{\frac{\sqrt{2}\alpha g}{\Delta\mp\Omega}\pm\frac{\Omega}{\sqrt{2}g\alpha^{*}}}\hat{\sigma}_{-}e^{i\omega_{c}t}+\f{\sqrt{2} g\alpha^{*}}{\Delta\mp\Omega}\hat{\sigma}_{+}e^{-i\omega_{c}t}+\f{\hat{\sigma}_{z}}{\sqrt 2}}~~.
  \label{eq:F_JC}
  \end{equation}
  where $N_{\pm}=\sqrt{2g^2|\alpha|^2}/\Omega$ is a normalization constant and $\Omega_{\mean{n}}=\Omega$.
 Their dynamics are of the form of Eq. \eqref{eq:eigoper} with corresponding eigenvalues $\theta_\pm\b t = \pm\Omega_{\mean{n}}t =\pm \sqrt{4|\alpha|^{2}g^{2}+\Delta^{2}}t$. The invariant operator reads
 \begin{equation}
     \hat{W}\b t =g\b{\alpha^*\hat{\sigma}_{-}e^{i\omega_{c}t}+\alpha \hat{\sigma}_{+}e^{-i\omega_{c}t}}+\f{\Delta}{{2}}\hat{\sigma}_{z}~.
     \label{eq:W_JC}
 \end{equation}
 Such an operator constitutes  a time-dependent constant of motion, since its expectation value remains constant under the free dynamics, Eq. \eqref{eq:JC_hamil}: $\mean{\hat{W}\b t}=\text{tr}\b{\hat{W}^H\b t \hat{\rho}_S\b 0}=\text{tr}\b{\hat{W}\b 0 \hat{\rho}_S\b 0}$.

%The eigenoperators of the control in the semi-classical limit evolve according to  $\hat{U}_C\hat{C}_k\hat{U}_C^\dagger=e^{-i\nu_k t}\hat{C}_k$ (Sec. \ref{sec:open_driven_system}), where  $\hat{U}_C$ is generated by the bosonic mode Hamiltonian $\hat{H}_C = \hbar \omega_c\b{\hat{a}^\dagger \hat{a}+\f{1}{2}}$. This relation implies that the eigenoperators of $\hat{U}_C$ are transition and invariant operators of $\hat{H}_C$,
%\begin{equation}
 %\{\hat{C}_k\}=\{\ket{n}\bra{m}\} ~~.
 %\label{eq:C_k_JC}
%\end{equation}

\subsection{Semi-classical limit for an open system}
\label{sec:JC_open}

The structure of the master equation for the combined system, including the qubit and the bosonic mode, can be obtained from the thermodynamical construction presented in Sec. \ref{sec:time_independent} and \ref{sec:semi_classical_limit}. The key ingredients of the resulting structure are
the primary-system and control semi-classical eigenoperators, which determine the structure of the semi-classical master equation. Inserting relations \eqref{eq:F_JC}, \eqref{eq:W_JC} 
into Eq. \eqref{eq:D_prime} leads to the reduced dynamics of a qubit, driven by a classical-field $\hat{V}\b t=\hbar g\b{\hat \sigma_{-}\alpha^*e^{i\omega_c t}+\hat \sigma_{+}\alpha e^{-i\omega_c t}}$. The dissipative part becomes
\begin{multline}
    {\cal{D}}^{s.c}\sb{\hat{\rho}_S\b t} =
    \sum_{k=\pm}\gamma_{k}\b t\b{  \hat{F}_{k}\hat{\rho}_S\b t\hat{F}_{k}^\dagger-\f{1}{2}\left\{\hat{F}_{k}^\dagger\hat{F}_{k},\hat{\rho}_S\b t \right\}}
    +\gamma_0\b t\b{\hat{W}\hat{\rho}_S\b t\hat{W}-\hat{\rho}_S\b t}~~,
    \label{eq:D_prime_JC}
\end{multline}
where $\hat{F}_\pm\equiv \hat{F}_\pm\b 0$ and $\hat{W}\equiv  \hat{W}\b 0$. Note, that generally corrections to the unitary dynamics, as the Lamb-shift, arise from the invariant term. If we absorb this into a an effective Lamb-shift Hamiltonian $\hat{H}_{LS}$, all the kinetic coefficients are positive under Markovian dynamics. The total master equation will then obtain the form 
${\cal L}^{s.c}\sb{\hat{\rho}_S\b t} = -\f{i}{\hbar}\sb{\hat{H}_S^{s.c}\b t+\hat{H}_{LS}^{s.c}\b t,\hat{\rho}_S\b t}+{\cal{D}}^{s.c}\sb{\hat{\rho}_S\b t}$. 
%where the Lindblad jump are given at initial time: $\hat{F}_k\equiv \hat{F}_k\b 0$ and $\hat{W}\equiv \hat{W}\b 0$. The kinetic coefficients in the master equation, $\Gamma_j$ and $\xi$, include a sum over all the terms $\gamma_k \text{tr}_C\b{\hat{C}_k^\dagger\hat{C}_k\hat{\rho}_C\b t}$ for which $\hat{F}_k=\hat{F}_j$, with $j=\pm$.

%\tg{We next demonstrate that the same dynamical equation can be obtained directly from the autonomous master equation Eq. \eqref{eq:D_int}, in the semi-classical limit. To derive the autonomous equation we first calculate the Lindblad jump operators of the autonomous equation  $\hat{G}_{k}$, which are the eigenoperators of the composite dynamics. These are easily obtained by separately diagonalizing each matrix $\hat{H}_S^{\b{n}}$, Eq. \eqref{eq:H_S^n} to obtain the eigenstates: $\ket{\psi_-^{\b{n}}}$ and $\ket{\psi_+^{\b{n}}}$ (given explicitly in Appendix \ref{apsec:JC_example}). Assuming the Bohr frequencies are non-degenerate, the transition (non-invariant) eigenoperators  $\{\hat{G}_k\}$are given by all the combinations of different eigenstates  $\{ \ket{\psi_\pm^{\b{n}}}\bra{\psi_\mp^{\b{m}}}\}$, and $\{\ket{\psi_\pm^{\b{n}}}\bra{\psi_\pm^{\b{m}}}\}$ for $n\neq m$. The invariant operators are linear combinations of the projection operators on the energy states $\{\hat{V}_j\}=\{\ket{\psi_\pm^{\b{j}}}\bra{\psi_\pm^{\b{j}}}\}$. Overall the autonomous master equation is given by
%\begin{equation}
 %  \tb{ ....}
%\end{equation}}

As expected, in the semi-classical limit tracing over the control degrees of freedom leads to a master equation, where the eigenoperators of $\hat{U}_S^{JC,s.c}\b t$ (with $n=\mean{n}$) constitute the Lindblad jump operator (see Eq. \eqref{eq:U_S^n} in Appendix \ref{apsec:convergence}).

\subsubsection{Kinetic coefficients and the physical significance of the global strict energy conservation }
The kinetic coefficients can be determined by employing a perturbative evaluation of the dynamical map \cite{dann2021non} or alternatively, in the Markovian regime one can supplement the axiomatic approach by a first principle derivation. 

We employ the construction of the NAME \cite{dann2018time} to determine the kinetic coefficients for the Jaynes-Cumming system coupled to a thermal bosonic bath. The derivation is valid in the Markovian regime, i.e, the typical timescale characterizing the decay of environmental correlation $\tau_E$ is much smaller then the relaxation timescale $\tau_R$ and the timescale associated with Jaynes-Cummings model internal dynamics $\tau_D$. In addition, the interaction with the environment is considered to be weak, i.e, $\tau_R\gg\tau_D$. Under these conditions the the following kinetic coefficients are obtained (see Appendix \ref{apsec:JC_open} for further details) 

\begin{equation}
    \gamma_0=\frac{2g^{2}|\alpha|^{2}}{\Omega^{2}}\b{J\b{|\omega_{c}|}\b{N\b{\omega_{c},T}+1}+J\b{|\omega_{c}|}N\b{\omega_{c},T}}
\nonumber
\end{equation}
\begin{equation}
    \gamma_{-}=s_{-}J\b{|\omega_{c}+\Omega|}\b{N\b{\omega_{c}+\Omega,T}+1}+s_{+}J\b{|\omega_{c}-\Omega|}N\b{\omega_{c}-\Omega,T}
    \label{eq:coeffic_sc}
\end{equation}
\begin{equation}
    \gamma_{+}=s_{-}J\b{|\omega_{c}-\Omega|}\b{N\b{\omega_{c}-\Omega,T}+1}+s_{+}J\b{|\omega_{c}+\Omega|}N\b{\omega_{c}+\Omega,T}
    \nonumber~~,
\end{equation}
where $s_{\pm}=\frac{\Delta\b{\Delta\pm\Omega\Delta}+2g^{2}|\alpha|^{2}}{4\Omega^{2}}$, $J$ is the bath's spectral density function and $N\b{\omega,T}$ is the Bose-Einstein distribution of a bath of temperature $T$.

Interestingly, the kinetic coefficients depend not only on the driving frequency $\omega_c\approx \omega_{eg}$, but also on the side-band frequencies $\omega_c\pm\Omega$. The specific dependence of the kinetic coefficients and the Lindblad jump operators can be understood by studying the transitions in the (autonomous) dressed basis $\{\ket{\psi_+^{\b n}},\ket{\psi_-^{\b n}}\}$, Eq. \eqref{eq:46}. The jump operators constitute of eigenoperators of the effective Hamiltonian $\hat{H}_{eff}=\f{\Delta}{2}\hat{\sigma_z}+\hbar g\b{\hat{\sigma}_{-}\alpha^{*}+\hat{\sigma}_{+}\alpha}$, with eigen energies $\pm=\pm\hbar\Omega/2$. 
As a result, the three terms of the master equation, proportional to $\gamma_-$ and $\gamma_+$, are related to  emission and absorption of an energy quanta $\hbar\Omega$.  Assuming a highly excited coherent state $\Omega_n\approx \Omega_\mean{n}=\Omega$ ($n\ra \mean{n}$) and the relevant transitions in the Jaynes-Cumming model are dominated by emission or absorption of photons of frequencies $\{\omega_c,\omega_c\pm\Omega,\Omega\}$. 
The first frequency is  a resonant transition which associated with $\gamma_0$, while an emission and absorption of a photon with a frequency $\omega_c+\Omega$ and $\omega_c-\Omega$ respectively, requires a transition from $\ket{\psi_+^{\b{\mean n}}}$ to $\ket{\psi_-^{\b{\mean n}}}$.  Therefore these transitions are associated with the the decay rate $\gamma_-$. Similarly the opposite transitions induce an excitation in the reduced description of the two-level system and are therefore incorporated in the excitation rate $\gamma_+$.   

The frequency dependence of the coefficients is connected to the so-called Mollow triplet or dynamical Stark effect \cite{burshtein1965kinetics,newstein1968spontaneous,mollow1969power,schuda1974observation} and is a strong indication that the our initial axioms hold effectively. Under resonance and sufficient light intensity (when the Rabi-frequency becomes comparable to the atomic linewidth) side-bands in the fluorescence spectrum start to appear. Along with the main peak, these three spectral peaks form the Mollow triplet.

A central assumption in our construction is that strict energy conservation between the device and the environment holds (postulate 4). However, a priori it is not clear whether this mathematical condition represents the actual physics. One can also imagine a possibility where the environment interact solely with the primary system, and the control is effectively isolated from the environmental effects (this is indeed the popular assumption in quantum optics). Nevertheless, the existence of the side-bands supports our initial choice.
The detection itself is a clear indication that the combined system (two-level system and harmonic model) emits to the environment (the lab and specifically to the detector) energy quanta of $\hbar\b{\omega_c\pm\Omega}$. In addition, the amounts of fluorescence depends on the density of states in the environment. If the density of state is small, the the transition of energy to the environment is suppressed or enhanced when the density is large \cite{kofman1994spontaneous,lambropoulos2000fundamental,PhysRevLett.107.193903}. This phenomena is captured by the dependence of the kinetic coefficients on the spectral density $J$ in the side-band frequencies. 
In contrast, the commonly employed local master equation, which induces changes only within the primary system energy levels, does does not capture this feature.

\subsection{Role of coherence in the Jaynes-Cummings model}

Coherence in the control system is manifested in two distinct ways in the autonomous and semi-classical descriptions.
In the autonomous description the coherence in the initial control state influences the kinetic coefficients of the reduced description (see also explanation in Sec. \ref{subsec:role_coherence}). Whereas, in the semi-classical description the initial coherence manifests itself within the Lindblad jump operators of the master equation. 

Specifically, in the JC model the jump operators of the (autonomous) master equation  are transition operators between the energy eigenstates $\{\ket{\psi_{\pm}^{\b n}}\}$, with
\begin{gather}
\ket{\psi_{+}^{\b n}}=\sin\b{\f{\alpha_{n}}2}\ket{g,n}+\cos\b{\f{\alpha_{n}}2}\ket{e,n-1}\nonumber\\\ket{\psi_{-}^{\b n}}=-\cos\b{\f{\alpha_{n}}2}\ket{g,n}+\sin\b{\f{\alpha_{n}}2}\ket{e,n-1}~~,
\label{eq:46}
\end{gather}
where $\alpha_n=\tan^{-1}\b{{\sqrt{n}\hbar g}}$.
The associated eigenenergies are $E_{\pm}^{\b n}=\pm \f{\hbar}2\sqrt{\Delta^{2}+4g^{2}n}+\hbar n\omega_{c}$, which imply that the device's Bohr frequencies are non-degenerate. This condition allows employing the construction of Sec. \ref{sec:time_independent} to write the general structure of the equation of motion which complies with the thermodynamic postulates (Sec. \ref{sec:framework_postulates}). The derived dynamical equation is of the form of the autonomous master equation, Eqs. \eqref{eq:L_decompos} and \eqref{eq:D}, with jump operators
$\{\ket{\psi_a^{\b{n}}}\bra{\psi_b^{\b{m}}}\}$ and $a,b=\pm$. 

When focusing on the primary-system dynamics, the superposition of bare control states in the device's eigenstates Eq. \eqref{eq:46}, leads to coherent contributions to the reduced master equation of the qubit. 
Following the general case in Sec. \eqref{subsec:role_coherence} 
reveals that the coherence in the control state is manifested by the contribution of `cross terms' $\mean{n|\hat{\rho}_C\b t|m}$ for $n\neq m$ to the kinetic coefficients of the autonomous reduced master equation. 

In the semi-classical master equation Eq. \eqref{eq:D_prime_JC}, the Lindblad jump operators are eigenoperators of the semi-classical propagator Eq. \eqref{eq:U_S^JC,s.c}, which in turn is generated by the semi-classical Hamiltonian $\hat{H}^{JC,s.c}_S=\text{tr}\b{\tilde{H}_D^{JC}\tilde{\rho}_C\b t}$ Eq. \eqref{eq:JC_hamil}. The dependence on $\hat{\rho}_C$  demonstrates that the identity of the jump operators is directly related to the initial control state. 

In the studied example, 
when the initial control state lacks coherence, the control parameter $\alpha$ vanishes and the semi-classical Hamiltonian becomes time-independent and is unaffected by the control. This behaviour is a general property of the semi-classical framework when the limit $g\ra 0$ is not compensated by large coherence in the control.   Hence, in the semi-classical description the time-dependence of the jump operators in the Schr\"odinger picture relies on large coherence in the control state.

\section{Discussion}
\label{sec:discussion}

The interrelation between thermodynamics and quantum mechanics suggests applying a combined approach towards the construction of quantum dynamical equations of motion. The main emphasis in the current analysis is on transient dynamics, where the obtained reduced equations of motion are explicitly time-dependent. To address this issue we employed two complementary frameworks to describe the same physical system, the autonomous and semi-classical descriptions. 
We showed that in the appropriate limit both methods lead to the same reduced dynamics for the primary-system. This is manifested by the fact that in this limit the Lindblad jump operators, which are the  eigenoperators of the free dynamics, coincide.

The main difference between the semi-classical and autonomous master equations (Eqs. \eqref{eq:D} and \eqref{eq:D_prime}) concerns the emergence of a transient character in the semi-classical framework. This is manifested by an explicit time-dependence in the kinetic and eigenoperators. 
The different temporal behaviour replaces the fixed point of the map by a time-dependent instantaneous attractor. Moreover, in contrast to the autonomous case, the semi-classical master equation mixes coherence and energy. The generated coherence can be traced back to the initial coherence in the control \cite{streltsov2017colloquium,lostaglio2015quantum}. Meaning that coherence is transferred from the control to the primary-system. As a consequence, if the control is initially in a stationary state, coherence and energy will evolve independently.
Therefore, the source of the transient character is the non-stationary state of the control. In this context there is a hidden assumption of a timescale separation between  the fast direct  influence of the environment on the primary-system and a slow indirect effect on the control.

In the semi-classical description the instantaneous attractor essentially serves as a moving target which the primary-system aspires to \cite{lindblad1975completely}. The target depends on the state of the control. Experiments in NV centers
\cite{holmstrom1997spin} 
have demonstrated the dependence of the dissipation on the control.
This property opens the doors to control the dissipative dynamics of the primary-system, which constitutes an open quantum system \cite{dann2019shortcut,dann2020fast}. 
{In addition, building upon the present study we have extended this framework to obtain a master equation which is applicable to non-Markovian scenarios \cite{dann2021non}.}

The thermodynamics of the {device} emerges from the condition of strict energy conservation which serves as isothermal partition between the system (including primary-system and control) and environment. As a consequence, heat is identified as the change of energy in the environment, which also coincides with the combined energy change in the primary-system and control.
If we additionally impose strict energy conservation between the primary-system and control, we can further partition the energy between the control and primary-system.
In this case, the change in energy in the control cannot be directly identified as work, since the energy transport may be accompanied by an increase in entropy \cite{scully2019laser}. 
Nevertheless, in the semi-classical limit, the primary-system and control state are separable. Under such a separation, the lack of correlations allows for energy transfer that leaves the entropy invariant. This property motivates the association of work with the change of energy in the {control}.

The definition of global correlations depend on the type of partition. The partitions are defined by the conserved quantities (or symmetries). Specifically, in the present study we imposed strict energy conservation between the device and the environment, Postulate 4, however other partitions are possible. 
A more comprehensive  description on this issue and the thermodynamics  will be given in a future study. 

{To summarize, in this study we presented an axiomatic construction of a master equation governing the dynamics of a driven open system. This approach generalizes previous perturbative derivations, based on the Davies construction. The framework illuminates the connection between the Lindblad jump operators of master equation to the dynamical symmetry \eqref{eq:maps_commute}. This symmetry is closely related to the specific isothermal partition between subsystems. In addition, the present construction supplies a clear connection between the complete quantum description (autonomous) and the pragmatic description, which employs a time-dependent drive (semi-classical). These two possible points of view, on the same physical system, enhance our insight on the interrelation between quantum mechanics and thermodynamics. }

%\begin{itemize}
 %   \item Discuss the dependence on the control state in the reduced master equation.
  %  \item Maybe compare to other time-dependent master equations
  %  \item Discuss the fixed point 
   % \item Discuss the thermodynamics of the obtained reduced dynamics
%    \item I-law and II-law
%\end{itemize}

%\section{DELETED MATERIAL}

\begin{acknowledgements}
 We thank Peter Salamon, James Nulton,
Erez Boukobza, Nina Megier, Christiane Koch, Josias Langbehn and Shimshon Kallush for insightful discussions. This research was supported by the Adams Fellowship  Program of the Israel Academy of Sciences and Humanities, the National Science Foundation under Grant No. NSF PHY-1748958 and The Israel Science Foundation Grant No.  526/21.
\end{acknowledgements}

\onecolumn\newpage
\appendix

\section{Propagators in different representations}
\label{secap:propagators}
The transition from the autonomous framework to the semi-classical description (Sec. \ref{sec:semi_classical_limit}) relied on Eq. \eqref{eq:relation1}, which constitutes a relation between the device propagator $\hat{U}_D\b{t,0}=e^{-i\hat{H}_{D} t/\hbar}$, the device propagator in the interaction picture $\tilde{U}_D\b{t,0}$ and the bare control propagator $\hat{U}_C\b{t,0}$. Here, we present a detailed derivation of this relation. 

Let $\ket{\psi}$ be the state of the device, which dynamics are governed by the Hamiltonian $\hat{H}_D$ Eq. \eqref{eq:autonomous_hamil}. The state in the interaction picture relative to $\hat{H}_C$ is defined as
\begin{equation}
    \ket{\tilde{\psi}\b t}=\hat{U}_C^\dagger\b{t,0}\ket{\psi\b t}~~.
    \label{eq:psi_tilde}
\end{equation}
The dynamics of $\ket{\tilde{\psi}}$ are governed by a Schr\"odinger equation
\begin{equation}
 i\hbar \pd{}{t}\ket{\tilde{\psi}}=\tilde{H}_D\b t\ket{\tilde{\psi}}~~,
 \label{eqap:int_psi}
\end{equation}
where 
\begin{equation}
    \tilde{H}_D\b t=\hat{H}_S+\hat{U}_C^\dagger\b{t,0}\hat{H}_{SC}\hat{U}_C\b{t,0}~~.
    \label{apeq:H_tilde}
\end{equation}
In turn, the propagator associated with Eq. \eqref{eqap:int_psi} reads 
\begin{equation}
    \tilde{U}_D\b{t,0} = {\cal T} \exp\b{-\f{i}{\hbar}\int_0^t\tilde{H}_D\b{\tau}d\tau}~~.
    \label{eqap:U_tilde}
\end{equation}
Utilizing Eqs. \eqref{eq:psi_tilde} and \eqref{eqap:U_tilde}, the dynamics of the device can then be expressed as
$\ket{{\psi}\b t}=\hat{U}_C\b{t,0} \tilde{U}_D\b{t,0}\ket{\psi\b{0}}$, which leads to 
\begin{equation}
    \hat{U}_D= \hat{U}_C \tilde{U}_D~~.
\end{equation}

\section{Time-translation symmetry in the semi-classical limit}
\label{apsec:time-translation}
We present a derivation showing that time-translation symmetry is satisfied in the semi-classical limit, i.e., the open system dynamical map commutes with isolated system map.
The relation is obtained by building upon the time-translation symmetry of the autonomous description (Eq. \eqref{eq:maps_commute}) as well as the procedure in Sec. \ref{sec:semi_classical_limit} used to derive the semi-classical limit.

We introduce the maps in the Heisenberg picture $\Lambda^{\ddagger}$ and ${\cal U}_D^{\ddagger}$. In this picture the time-translation symmetry obtains the form
\begin{equation}
    \Lambda^{\ddagger}\sb{{\cal U}_{D}^{\ddagger}\sb{\hat O_{D}}}={\cal U}_{D}^{\ddagger}\sb{\Lambda^{\ddagger}\sb{\hat O_{D}}}~~,
    \label{apeq:22}
\end{equation}
where $\hat{O}_D$ is a general device operator.

By expressing the partial trace on the environment in terms of the eigenstates of $\hat{\rho}_E\b 0$: $\{\ket{\chi_i}\}$, Eq. \eqref{apeq:22} becomes (cf. Ref. \cite{dann2020open})
\begin{equation}
    \sum_{i}\bra{\chi_{i}}\hat U^{\dagger}\hat U_{D}^{\dagger}\hat O_{D}\hat U_{D}\hat U\ket{\chi_{i}}=\sum_{i}\bra{\chi_{i}}\hat U_{D}^{\dagger}\hat U^{\dagger}\hat O_{D}\hat U\hat U_{D}\ket{\chi_{i}}~~.
    \label{apeq:s1}
\end{equation}
In the semi-classical limit, the device's time-evolution operator can be decomposed as $ \hat{U}_D \cong \hat{U}_C{\otimes}\hat{U}_S^{s.c}$, Eq. \eqref{eq:sc_limit_prop}.
A similar decomposition can be obtained for $\hat{U}$, in order to derive this we repeat the procedure of Sec. \eqref{sec:semi_classical_limit} for the total system. The composite time-evolution operator can be equivalently written as $\hat{U}=\hat{U}_C\tilde{U}$, where the overscript tilde designates operators in the interaction picture relative to the bare control dynamics. In analogy with Eqs. \eqref{eq:relation1} and \eqref{eq:interaction_picture_Hamil}, $\tilde{U}$ is generated by the interaction Hamiltonian $\tilde{H}\b t=\hat{H}_{S}+
\hat U_{C}^{\dagger}\b{t,0}\hat H_{SC}\hat U_{C}\b{t,0}+\hat H_{SE}+\hat H_{E}$.
Following the semi-classical procedure we integrate over the Liouville-von Neumann equation in the interaction picture and trace over the control, this leads to an analogous relation to Eq. \eqref{eq:15rho}
\begin{equation}
    \hat \rho_{SE}\b t=-\f i{\hbar}\int_0^t {\text{tr}_{C}\b{\sb{\hat H_{S}+\hat H_{SE}+\hat H_{E}+\sum_{i}\hat S\otimes\tilde{C}_{i}\b{\tau},\tilde{\rho}_{SE}\b{\tau}}}d\tau}
    \label{eqap:111}
\end{equation}
where $\hat{\rho}_{SE}$ represents the combined state of the primary system and environment (excluding the control). 
Next, we employ the semi-classical conditions,
$\hat \rho\b t=\hat \rho_{SE}\b t\otimes\hat \rho_{C}\b t$  and $\hat \rho_{C}\b t=\hat U_{C}\b{t}\hat \rho_{C}\b 0\hat U_{C}^{\dagger}\b{t}$. Substituting these relation into Eq. \eqref{eqap:111} leads to
\begin{equation}
\f d{dt}\hat \rho_{SE}\b t -\f i{\hbar}\int_0^t {\sb{\hat{H}^{s.c}\b{\tau},\hat \rho_{SE}\b{\tau}}d\tau}~~,
\end{equation}
where $\hat{H}^{s.c}\b t=\text{tr}\b{\tilde{H}\tilde{\rho}_C \b t}$. This equation implies that in the semi-classical limit the time-evolution operator can be decomposed to a propagator of the control and a semi-classical propagator on the primary-system and environment state
\begin{equation}
    \hat U\simeq \hat U_{C}\otimes\hat U_{SE}^{s.c}~~,
    \label{apeq:112}
\end{equation}
where $\hat U_{SE}^{s.c}$ satisfies the Schr\"odinger equation with respect to the semi-classical composite Hamiltonian $\hat{H}^{s.c}\b t$.

Finally, we consider a primary-system operator $\hat{O}_D = \hat{O}_S\otimes\hat{I}_C$, substitute relations \eqref{apeq:112} and \eqref{eq:sc_limit_prop} into Eq. \eqref{apeq:s1} and trace over the control degrees of freedom. The left hand side then becomes
\begin{multline}
    \text{tr}_{C,E}\b{\hat U^{\dagger}\hat U_{D}^{\dagger}\hat O_{S}\hat U_{D}\hat U}=\text{tr}_{C,E}\b{\hat U_{SE}^{s.c\dagger}\hat U_{C}^{\dagger}\hat U_{S}^{s.c\dagger}\hat U_{C}^{\dagger}\hat O_{S}\hat U_{C}\hat U_{S}^{s.c}\hat U_{C}\hat U_{SE}^{s.c}}\\=\text{tr}_{C,E}\b{\hat U_{SE}^{s.c\dagger}\hat U_{S}^{s.c\dagger}\hat O_{S}\hat U_{S}^{s.c}U_{\hat SE}^{s.c}}~~,
\end{multline}
while the right hand side is given by 
\begin{equation}
\text{tr}_{C,E}\b{\hat U_{D}^{\dagger}\hat U^{\dagger}\hat O_{D}\hat U\hat U_{D}}=\text{tr}_{C,E}\b{\hat U_{S}^{s.c\dagger}\hat U_{SE}^{s.c\dagger}\hat O_{S}\hat U_{SE}^{s.c}\hat U_{S}^{s.c}}~~.
\label{apeq:64}
\end{equation}

We now identify the semi-classical open system dynamical map $$\Lambda^{s.c}\sb{\bullet}=\text{tr}_{E}\b{U_{SE}^{s.c}\bullet U_{SE}^{s.c\dagger}}
~~,$$ and express Eq. \eqref{apeq:64} as
\begin{equation}
    \Lambda^{s.c\ddagger}\sb{{\cal U}_{S}^{s.c\ddagger}\sb{O_{S}}}={\cal U}_{S}^{s.c\ddagger}\sb{\Lambda^{s.c\ddagger}\sb{O_{S}}}~~,
\end{equation}
or equivalently in the Sch\"ordinger picture 
\begin{equation}
    \Lambda^{s.c}\circ{\cal U}_{S}^{s.c}={\cal U}_{S}^{s.c}\circ\Lambda^{s.c}~~.
\end{equation}

\section{Instantaneous attractor}
\label{apsec:i.a.}
In the following section we explicitly derive the action of the dynamical map on the instantaneous attractor $\tilde{\rho}_S^{i.a}$. Deriving in detail the relation
\begin{equation}
\widetilde{\cal{D}}_S\sb{\tilde{\rho}_S^{i.a}}=0~~.
\label{apeq:D_S}
\end{equation}

The commutation relations of the Lindblad jump operators and the effective Hamiltonian \eqref{eq:comutation_eff}, and the Bake-Campell-Housdorff formula lead to 
\begin{equation}
    e^{-\bar{H}_S}\hat{F}_k = e^{\delta_k}\hat{F}_k e^{-\bar{H}_S}~~.
\end{equation}
In addition, the generator can be composed to a sum over independent channels $\widetilde{\cal{D}}_S=\sum_k \widetilde{\cal{D}}_S^{\b{k}}$, where $\widetilde{\cal{D}}_S^{\b{k}}$ is given by  (Eq. \eqref{eq:eq28})
\begin{equation}
    \widetilde{\cal{D}}_S^{\b{k}}\sb{\bullet} = \Gamma_k\b{  \hat{F}_{k}\bullet\hat{F}_{k}^\dagger-\f{1}{2}\left\{\hat{F}_{k}^\dagger\hat{F}_{k},\bullet \right\}}+\Gamma_{-k}\b{  \hat{F}_{k}^\dagger\bullet\hat{F}_{k}-\f{1}{2}\left\{\hat{F}_{k}\hat{F}_{k}^\dagger,\bullet \right\}}~~.
    \label{apeq:28}
\end{equation}
This allows restricting the analysis to a single channel $k$.  
The action of the first term in the $k$'th channel is 
\begin{multline}
    \Gamma_k\sb{\hat{F}_{k}e^{-\bar{H}_{S}}\hat{F}_{k}^{\dagger}-\f 12\{\hat{F}_{k}^{\dagger}\hat{F}_{k},e^{-\bar{H}_{S}}\}}\\=\Gamma_k\sb{e^{-\delta_{k}}\hat{F}_{k}\hat{F}_{k}^{\dagger}e^{-\bar{H}_{S}}-\f 12\b{\hat{F}_{k}^{\dagger}\hat{F}_{k}e^{-\bar{H}_{S}}+e^{-\delta_{k}}\hat{F}_{k}^{\dagger}e^{-\bar{H}_{S}}\hat{F}_{k}}}\\=\Gamma_k\sb{e^{-\delta_{k}}\hat{F}_{k}\hat{F}_{k}^{\dagger}-\hat{F}_{k}^{\dagger}\hat{F}_{k}}e^{-\bar{H}_{S}}
\end{multline}
And similarly for the second term in Eq. \eqref{apeq:28}
\begin{equation}
    \Gamma_{-k}\sb{\hat{F}_{k}^\dagger e^{-\bar{H}_{S}}\hat{F}_{k}-\f 12\{\hat{F}_{k}\hat{F}_{k}^{\dagger},e^{-\bar{H}_{S}}\}}=\Gamma_{-k}\sb{e^{\delta_{k}}\hat{F}_{k}^\dagger\hat{F}_{k}-\hat{F}_{k}\hat{F}_{k}^{\dagger}}e^{-\bar{H}_{S}}~~.
\end{equation}
For $\Gamma_{k}/\Gamma_{-k}=e^{\delta_k}$, the two terms cancel each other, leading to $\widetilde{\cal D}_S^{\b k}\sb{\tilde{\rho}_S^{i.a}}=0$. A similar result is obtained for all the channels, determining the identity of the $\delta_k$ parameters.  Summing over the channels we obtain the desired expression
\begin{equation}
 \widetilde{\cal D}_S\sb{\tilde{\rho}_S^{i.a}}=0~~.    
\end{equation}

\section{Liouville space and the unitarity of the Liouville propagator}
\label{apsec:unitity_of_U}

Consider a compact system  with an associated (wave-function) Hilbert space of dimension $M\in\mathbb{N}$. The equivalent description of the quantum state in Liouville space (known also as Hilbert-Schmidt space) includes defining an operator basis $\{\hat{X}\}$, which spans the Hilbert space of system operators. 
A general operator is then expressed in terms of the basis operators $\hat{O}=\sum_i \hat{c_i}\hat{X}_i$, with $c_i =\b{\hat{X_i},\hat{O}} \equiv\text{tr}\b{\hat{X}_i\hat{O}}$, which defines the corresponding vector in Liouville space: $\v{o}=\{c_1,\dots,c_N\}^T$, where $N=M^2$ is dimension of the operators basis. A superoperator $\cal S$ operating on operator $\hat{O}$, is represented by an $N$ by $N$ matrix with elements ${\cal S}_{ij} = \text{tr}\b{\hat{X}_i^\dagger{\cal S}\sb{{\hat{X}_j}}}$.

\paragraph*{Vec-ing}
The vec-ing procedure refers to a certain choice of an operator basis. This operator basis flattens an operator $\hat{O}$ into an $N=M^2$ dimensional vector. The mapping is defined such that the $\b{a,b}$ entry of $\hat{O}$ is mapped to the $\b{b-1}M+a$ entry of the Liouville vector $\v{o}$. If $\hat{O}$ is represented by the wave-function basis $\{\ket{n}\}$, where $M\geq n\in\mathbb{N}$ then the vec-ing procedure is a representation of operators in terms of
the basis 
$\{\ket{0}\bra{0},\ket{1}\bra{0},\dots,\ket{m}\bra{0},\ket{0}\bra{1},\dots,\ket{m}\bra{m}\}$. 
For such a choice of basis, the superoperator $\hat{A}\bullet\hat{B}$ is mapped to $\hat{B}^T\otimes\hat{A}$.

\paragraph*{Unitarity of the Liouville propagator}
Utilizing the vec-ing procedure we find that ${\cal{H}}=\sb{\hat{H},\hat{\rho}}$ is mapped to the matrix $\overleftrightarrow{\cal{H}}=\hat{I}\otimes\hat{H}-\hat{H}^T\otimes \hat{I}$. When $\hat{H}$ is an Hermitian operator, as in the case of the Heisenberg equation,  $\overleftrightarrow{\cal{H}}$ is Hermitian 
\begin{equation}
    \overleftrightarrow{\cal{H}}^\dagger =\hat{I}\otimes\hat{H}^\dagger -\hat{H}^*\otimes \hat{I}=
    \hat{I}\otimes\hat{H} -\hat{H}^T\otimes \hat{I}=\overleftrightarrow{\cal H}~.
\end{equation}
which implies that the propagator in Liouiville space $\overleftrightarrow{\cal U}^{\ddagger}$, defined by $\pd{}{t}\overleftrightarrow{\cal U}=i{\overleftrightarrow{\cal{H}}}\overleftrightarrow{\cal U}$ is unitary.

\section{Convergence to the semi-classical limit in the Jaynes-Cummings Hamiltonian}
\label{apsec:convergence}

The semi-classical description of the Jaynes-Cummings model is obtained by considering a bosonic mode for which $\mean{n}\gg 1$, while keeping $g\sqrt{n}$ constant and moderate compared to the qubit's internal frequency. To obtain this limit rigorously we study the autonomous system dynamics and derive the dynamics of $\rho_S\b t$ in this regime.

%We start by analysing the dynamics in the interaction picture relative to the control. In this picture the propagator 
%is given by $\tilde{U}_S=\hat{U}_C^\dagger\hat{U}_S = \sum_n \hat{U}_C^\dagger \hat{U}_S^{\b{n}}$, and is generated by the Hamiltonian in the interaction picture, Eq. \eqref{eq:interaction_picture_Hamil}, see Sec. \ref{sec:JC_model} for more details. 
%Utilizing Eq. \eqref{eq:U_S^n} one can express the  propagator of the subspace $n$ as
%\begin{multline}
 %   \tilde{U}_S^{\b n}=\\
  %  =\left(\begin{array}{cc}
%e^{i\omega_{c}t/2}\left(\text{c}_n+\frac{i\Delta\text{s}_n}{\Omega_{n}}\right) & -ie^{i\omega_{c}t/2}\f{\Omega\sqrt{n}}{\Omega_{n}}\text{s}_n\\
%-ie^{-i\omega_{c}t/2}\f{\Omega\sqrt{n}}{\Omega_{n}}\text{s}_n & e^{-i\omega_{c}t/2}\left(\text{c}_n-\frac{i\Delta\text{s}_n}{\Omega_{n}}\right)
%\end{array}\right)~~,
%\end{multline}
%with $\text{c}_n=\cos\left(\frac{\Omega_{n}t}{2}\right)$, $\text{s}_n=\sin\left(\frac{\Omega_{n}t}{2}\right)$ and $\Omega_n=\sqrt{\Delta^{2}+n\Omega^{2}}$.

For an initial control state $\hat{\rho}_C=\ket{\alpha}\bra{\alpha}$
the reduced state of the qubit becomes
\begin{equation}
   \hat{\rho}_S\b t=\text{tr}_{C}\b{\hat{U}_D^{JC}\b{\hat{\rho}_{S}\b 0\otimes\hat{\rho}_{C}}\hat{U}^{JC \dagger }_D} =\sum_{m=0}^{\infty}\hat{\chi}_{m}\b t\rho_{S}{\b 0}\hat{\chi}_{m}^\dagger\b t~~,
   \label{apeq:kraus_form}
\end{equation}
with Kraus operators
\begin{equation}
  \hat{\chi}_{m} \b t\equiv \bra{m}  \hat{U}_S\ket{\alpha\b t} =\sum_{n=0}^{\infty}  \bra{m}  \hat{U}_D^{\b{n}}\ket{\alpha\b t}~~.
    \label{eq:chi_m}
\end{equation}
Here, $\hat{U}_D^{\b{n}}$ is the propagator of the $n$'th subspace of $\hat{H}_D^{\b{n}}$,  Eq. \eqref{eq:H_S^n} and $\hat{U}_D^{JC}$ is generated by the Jaynes-Cumming Hamiltonian $\hat{H}_D^{JC}$, Eq. \eqref{eq:JC_hamil}. Written explicitly in terms of the basis $\{\ket{g,n},\ket{e,n-1}\}$, the subspace propagator becomes  
\begin{multline}
    \hat{U}_D^{\b{n}}\b{t,0}=\\
    \left(\begin{array}{cc}
e^{-in\omega_{c}t}\left(\text{c}_{{n}}+\frac{i\Delta\text{s}_{{n}}}{\Omega_{{\mean{n}}}}\right) & -ie^{-in\omega_{c}t}\f{2 g\sqrt{{{n}}}}{\Omega_{{{n}}}}\text{s}_n\\
-ie^{-in\omega_{c}t}\f{2 g\sqrt{{{n}}}}{\Omega_{{{n}}}}\text{s}_n & e^{-int\omega_{c}t}\left(\text{c}_{{n}}-\frac{i\Delta\text{s}_{{n}}}{\Omega_{{{n}}}}\right)
\end{array}\right)~~,
\label{eq:U_S^n}
\end{multline}
with $\text{c}_n=\cos\left(\frac{\Omega_{n}t}{2}\right)$, $\text{s}_n=\sin\left(\frac{\Omega_{n}t}{2}\right)$ and $\Omega_n=\sqrt{\Delta^{2}+4g^{2}n}$.
Utilizing Eq. \eqref{eq:U_S^n} and \eqref{eq:chi_m} the Kraus operators can be expressed in a matrix form  
\begin{multline}
    \hat{\chi}_m\b t=e^{-|\alpha|^{2}/2}\times\\\left(\begin{array}{cc}
{c_{m}+i\f{\Delta s_{m}}{\Omega_{m}}}\f{\alpha^{m}}{\sqrt{m!}} & -i\f{2g\sqrt{m}}{\Omega_{m}}s_{m}\f{\alpha^{m-1}}{\sqrt{\b{m-1}!}}\\
-i\f{2g\sqrt{m+1}}{\Omega_{m+1}}s_{m+1}\f{\alpha^{m+1}}{\sqrt{m+1!}} & {c_{m}-i\f{\Delta s_{n}}{\Omega_{n}}}\f{\alpha^{m-1}}{\sqrt{m!}}
\end{array}\right)~~,
\end{multline}
where the chosen basis is $\{\ket{g},\ket{e}\}$

In the semi-classical limit the major contribution is associated with $m\approx|\alpha|^2\gg1$. In this regime $\sqrt{\b{m\pm1}!}=\sqrt{m!}+O\b{\f{1}{m!}}$, $\text{s}_{n\pm1}=\text{s}_n+ O\b{|\alpha|^{-2}}$, and  $\text{c}_{n\pm1}=\text{c}_n+ O\b{|\alpha|^{-2}}$, leading to
\begin{multline}
    \hat{\chi}_m\b t\approx\f{\alpha^{m}}{\sqrt{m!}}e^{-|\alpha|^{2}/2}\times\\\left(\begin{array}{cc}
c_{m}+i\f{\Delta s_{m}}{\Omega_{m}} & -i\f{2g\sqrt{m}}{\Omega_{m}}s_{m}\\
-i\f{2g\sqrt{m}}{\Omega_{m}}s_{m} & {c_{m}-i\f{\Delta s_{m}}{\Omega_{m}}}
\end{array}\right)~~.
\label{apeq:chi_m}
\end{multline}

To evaluate the Kraus form, Eq. \eqref{apeq:kraus_form}, in the semi-classical limit, it is first useful to calculated the asymptotic behaviour of the sum $e^{-|\alpha|^{2}}\sum_{m=0}^{\infty}\f{|\alpha|^{2m}}{m!}f\b m$, where $f$ is  an analytical 
function of $m$. Expressing $f\b{m}$ in terms of a Maclaurin series, $f\b{m}=\sum_j c_j m^j$ with $c_j = d^jf/dm^j|_{m=0}$ leads to 
\begin{multline}
    e^{-|\alpha|^{2}}\sum_{m=0}^{\infty}\f{|\alpha|^{2m}}{m!}f\b m\\=e^{-|\alpha|^{2}}\sum_{m=0}^{\infty}\f{|\alpha|^{2m}}{m!}\sum_{j}c_{j}m^{j}\\=e^{-|\alpha|^{2}}\sum_{j}c_{j}\sum_{m=0}^{\infty}\f{|\alpha|^{2m}}{m!}m^{j}\\=\sum_{j}c_{j}T_{j}\b{|\alpha|^{2}}~~,
    \label{apeq:derivation}
\end{multline}
where $T_{j}\b x=e^{-x}\sum^\infty_{k=0} \f{k^n x^k}{k!}$ are the Touchard polynomials. For $x\in \mathbb{R}$ these polynomials satisfy the following relation asymptotically ($|x|\ra\infty$) \cite{paris2016asymptotics}
\begin{equation}
    x^{-j} T_j\b{x}=1+{j\b{j-1}}\b{\f{1}{2x}+O\b{\f{1}{x^2}}}~~.
\end{equation}
Thus, in the classical limit, for which $|\alpha|^2\gg1$, the Touchard polynomials in Eq. \eqref{apeq:derivation} become 
\begin{equation}
    T_{j}\b{|\alpha|^2}=|\alpha|^{2j}\b{1+O\b{\f{1}{|\alpha|^2}}}~~.
    \label{apeq:T_j}
\end{equation}
Substituting \eqref{apeq:T_j} into Eq. \eqref{apeq:derivation} gives the desired expression
\begin{equation}
    e^{-|\alpha|^{2}}\sum_{m=0}^{\infty}\f{|\alpha|^{2m}}{m!}f\b m\cong f\b{|\alpha|^2}~~.\\
    \label{apeq:relation}
\end{equation}
Gathering Eqs. \eqref{apeq:kraus_form}, \eqref{apeq:chi_m} and  \eqref{apeq:relation} we obtain the qubit's dynamics in the semi-classical limit
\begin{equation}
    \hat{\rho}_S\b t \cong \hat{U}_S^{JC,s.c} \b{t,0}\hat{\rho}_S\b 0 \hat{U}_S^{JC,s.c\dagger}\b{t,0}~~,
\end{equation}
with the semi-classical propagator
\begin{equation}
     \hat{U}_S^{JC,s.c}=\left(\begin{array}{cc}
{c_{\mean n}+i\f{\Delta s_{\mean n}}{\Omega_{\mean n}}} & -i\f{2g\sqrt{\mean n}}{\Omega_{\mean n}}s_{\mean n}\\
-i\f{2g\sqrt{\mean n}}{\Omega_{\mean n}}s_{\mean n} & {c_{n}-i\f{\Delta s_{\mean n}}{\Omega_{\mean n}}}
\end{array}\right)~~,
\label{apeq:U_S}
\end{equation}
where $\mean{n}=\bra{\alpha}\hat{a}^\dagger\hat{a}\ket{\alpha}$ is the mean photon number of the coherent state.

\section{Comparison of the autonomous solution in the semi-classical limit with the semi-classical solution (Rabi model)}
\label{apsec:comparison}
The semi-classical Hamiltonian is obtained by tracing over the environment degrees of freedom, as defined bellow Eq. \eqref{eq:U_sc}. In the Jaynes-Cummings model this definition leads to Eq. \eqref{eq:JC_hamil}:
\begin{equation}
        \hat{H}_{S}^{JC,s.c}\b t = \f{\hbar\omega_{eg}}2\hat \sigma_{z}+\hbar g\b{\hat \sigma_{-}\alpha^*e^{i\omega_c t}+\hat \sigma_{+}\alpha e^{-i\omega_c t}}~~.
    \label{apeq:JC_hamil}
\end{equation}
Performing a rotation by an angle $\varphi=-\text{arg}\b{\alpha}$ around the $z$ axis, we obtain an equivalent representation of the Hamiltonian
\begin{multline}
    \hat{H}_{S}^{JC,s.c}\ra\hat{R}_z^\dagger\b{\varphi}\hat{H}_S^{JC,s.c}\hat{R}_z\b{\varphi}\\
     = \f{\hbar\omega_{eg}}2\hat \sigma_{z}+\hbar g|\alpha|\b{\hat \sigma_{-}e^{i\omega_c t}+\hat \sigma_{+} e^{-i\omega_c t}}~~,
    \label{apeq:JC_hamil}
\end{multline}
where $\hat{R}_z\b{\varphi}=e^{-i\sigma \varphi/2}$.
The semi-classical dynamics can be obtained by transforming to an interaction (rotating) picture, defined by $\ket{\psi_{V}}=V^{\dagger}\ket{\psi}$ with $V=e^{-i\omega_{c}\sigma_{z}t/2}$, and diagonalizing the resulting Hamiltonian. This procedure leads to the  propagator $\hat{U}_S^{JC,s.c}$ Eq. \eqref{eq:U_sc}, which is equivalent to the propagator in Eq. \eqref{apeq:U_S}, that was obtained from the autonomous dynamics by taking the semi-classical limit.

\section{First principle derivation of the of Jaynes-Cumming master equation}
\label{apsec:JC_open}
We present a first principle derivation of Eq. \eqref{eq:coeffic_sc}, the derivation supplements the exact operatorial structure of Eq. \eqref{eq:D_prime_JC} and allows evaluating the kinetic coefficients in a certain physical regime.

Assuming weak coupling between and Markovian dynamics, the Bohr Markov approximation leads to the Quantum Markovian master equation \cite{breuer2002theory}.
\begin{equation}
    \f{d\tilde{\rho}_{S}\b t}{dt}=-\f{1}{\hbar^2}\int_ 0^{\infty}{ds\,\text{tr}_{E}\b{\sb{\tilde{H}_{I}\b t,\sb{\tilde{H}_{I}\b{t-s},\tilde{\rho}_{S}\b t\otimes\hat{\rho}_{E}}}}}~~,
    \label{eqap:QME}
\end{equation}
where $\tilde{\rho}_S\b t$ and $\tilde{H}_I$ are the qubits state and interaction term, in the interaction representation relative to the semi-classical free dynamics $\hat{H}_S^{s.c}\b t+\hat{H}_E$. In addition, due to the Markovian nature of the dynamics the primary system and environment can be effectively described in terms of a separable state, where the environment remains in a stationary state.

We consider a bosonic bath of temperature $T$ and an interaction of the form
\begin{equation}
    \hat{H}_{I} = \hat{\sigma}_x \otimes \hat{E}
    \label{eqap:H_I}
\end{equation}
where $\hat{E} =\sum_k g_k\b{\hat{b}_k+\hat{b}_k^\dagger }$ is the environment interaction term.
This interaction does satisfy the strict energy conservation condition of postulate 4, nevertheless, the derivation will lead a master equation which complies with the operatorial structure and dynamical symmetry dictated by postulate 4. The apparent  incongruity can be as an effective interaction under a coarse-graining in time. That is, for sufficiently long timescales  the effect of the ``non-energy conserving'' terms, which do not satisfy the strict energy conservation, average out and the interface energy between the device and environment is effectively constant \cite{winczewski2021bypassing}. 

Next, we expand the system interaction term in terms of the eigenoperators, leading to 
\begin{multline}
    \tilde{\sigma}_{x}\b t=\hat{U}_{S}^{s.c\dagger}\b t\hat{\sigma}_{x}\hat{U}_{S}^{s.c}\b t=\frac{\sqrt{2}g}{\Omega}\b{\alpha\hat{F}_{0}e^{-it{{\omega}_c}}+\alpha^{*}\hat F_{0}e^{it{{\omega}_c}}}\\+s_{+}^{\b 1}\hat F_{+}e^{-i\b{\omega_{c}-\Omega}t}+s_{+}^{\b 2}\hat F_{+}e^{i\b{\omega_{c}+\Omega}t}\\+s_{-}^{\b 1}\hat F_{-}e^{-i\b{\omega_{c}+\Omega}t}+s_{-}^{\b 2}\hat F_{-}e^{i\b{\omega_{c}-\Omega}t}~~,
\end{multline}
where $\hat{F}_0\equiv \hat{F}_0\b 0$,$\hat{F}_\pm\equiv \hat{F}_\pm\b 0$, and 
\begin{equation}
    s_{\pm}^{\b 1}=\pm\f{2g^{2}|\alpha|\alpha^{*}}{\Omega\b{\Omega\pm\Delta}}\,\,\,\,\,;\,\,\,\,\,s_{\pm}^{\b 2}=\b{\mp\f 1{|\alpha|}\pm\f{2g^{2}|\alpha|}{\Omega\b{\Omega\pm\Delta}}}\alpha~~.
\end{equation}
Substituting Eq. \eqref{eqap:H_I} into Eq. \eqref{eqap:QME} and performing the secular approximation (terminating fast oscillating terms, valid under the condition that $\tau_R\gg\tau_S$) leads to the following master equation
\begin{multline}
    \f d{dt}\tilde{\rho}_{S}\b t=\gamma_{-}\b{\hat F_{-}\tilde{\rho}_{S}\b t\hat F_{+}-\f 12\{\hat F_{+}\hat F_{-},\tilde{\rho}_{S}\b t\}}\\+\gamma_{+}\b{\hat F_{+}\tilde{\rho}_{S}\b t\hat F_{-}-\f 12\{\hat F_{-}\hat F_{+},\tilde{\rho}_{S}\b t\}}\\+\gamma_{0}\b{\hat F_{0}\tilde{\rho}_{S}\b t\hat F_{0}-\tilde{\rho}_{S}\b t}~~.
\end{multline}
The kinetic coefficients are given by 

\begin{gather}
    \gamma_{0}=k_0\b{\Gamma\b{\omega_{c}}+\Gamma\b{-\omega_{c}}} \nonumber\\\gamma_{-}=k_{-}\Gamma\b{\omega_{c}+\Omega}+k_{+}\Gamma\b{\omega_{c}-\Omega}\\\gamma_{+}=k_{-}\Gamma\b{\omega_{c}-\Omega}+k_{+}\Gamma\b{\omega_{c}+\Omega} \nonumber~~,
\end{gather}
where
\begin{equation}
    \Gamma\b{\nu}=\int_0^{\infty}{ds\,}e^{i\nu s}\mean{\hat{E}\b t\hat{E}\b{t-s}}
    \label{eqap:Gamma}
\end{equation}
is the one-sided Fourier transform of the environment correlation function and 
\begin{equation}
    k_0 =\frac{2g^{2}|\nu|^{2}}{\Omega^{2}}~~~~;~~~~k_{\pm}=\frac{\Delta\b{\Delta\pm\Omega\Delta}+2g^{2}|\nu|^{2}}{4\Omega^{2}}
\end{equation}
are constant coefficients.
Following the derivation in Ref. \cite{breuer2002theory} Sec. 3 Eq. \eqref{eqap:Gamma} becomes 
\begin{equation}
    \Gamma\b{\nu} = |g_{k\b{\nu}}|^2\Phi\b{\nu}\b{N\b{\nu,T}+1}~~,
\end{equation}
where $k\b{\nu}$ is designates the environment mode which oscillates at frequency $\nu$, $\Phi\b{\nu}$ is the field density of state at frequency $\nu$, (for example for a three-dimensional field $\Phi\b{\nu}\propto \nu^2$) and $N\b{\nu,T} = \f{1}{e^{\hbar\nu/k_BT}-1}$.

\bibliographystyle{abbrvnat}
%\bibliographystyle{ksfh_nat}
%\bibliographystyle{plainnat}

%\bibliographystyle{vancouver}

%\bibliography{references.bib}

\end{document}